\documentclass[12pt,notitlepage,a4paper]{article}

\pdfoutput=1

\usepackage{a4wide}
\usepackage{epsfig}
\usepackage{color,graphicx}
\usepackage{cite}
\usepackage{amssymb,amsmath}
\usepackage{jheppub}

\newcommand{\be}{\begin{equation}}
\newcommand{\ee}{\end{equation}}
\newcommand{\bea}{\begin{eqnarray}}
\newcommand{\eea}{\end{eqnarray}}

\usepackage[dvipsnames]{xcolor}

\usepackage{graphics, color,tikz,mathtools, colortbl, xcolor,enumerate, tikz-3dplot, slashed,bm}

\usetikzlibrary{positioning}
\usetikzlibrary{chains}
\usetikzlibrary{arrows, arrows.meta ,fit,decorations.pathreplacing,decorations.pathmorphing}
\tikzstyle{every picture}+=[remember picture]
\tikzstyle{na} = [baseline]

\usetikzlibrary{arrows, decorations.markings, calc, fadings, decorations.pathreplacing, patterns, decorations.pathmorphing, positioning}
\tikzset{>={Latex[width=1.5mm,length=1.5mm]}}
\usetikzlibrary{graphs,graphs.standard,quotes}

\tikzset{->-/.style={decoration={
  markings,
  mark=at position #1 with {\arrow{>}}},postaction={decorate}}}    
\tikzset{-<-/.style={decoration={
  markings,
  mark=at position #1 with {\arrow{<}}},postaction={decorate}}}  
  
\newcommand{\nn}{\nonumber}
\def\ga{\alpha}
\def\gb{\beta}
\def\gc{\gamma}

\def\gd{\delta}

\begin{document}

\begin{center}  

\vskip 2cm 

\centerline{\Large {\bf\boldmath Trinions for the $3d$ compactification}}
\centerline{\Large {\bf\boldmath of the $5d$ rank 1 $E_{N_f+1}$ SCFTs}}

\vskip 1cm

\renewcommand{\thefootnote}{\fnsymbol{footnote}}

   \centerline{
    {\large \bf Matteo Sacchi${}^{a}$} \footnote{matteo.sacchi@maths.ox.ac.uk}{\large \bf , Orr Sela${}^{b}$} \footnote{osela@physics.ucla.edu} {\large \bf and Gabi Zafrir${}^{c,d}$} \footnote{gzafrir@scgp.stonybrook.edu}}
      
\vspace{1cm}
\centerline{{\it ${}^a$ Mathematical Institute, University of Oxford, Andrew-Wiles Building, Woodstock Road,}}
\centerline{{\it Oxford, OX2 6GG, United Kingdom}}
\centerline{{\it ${}^b$ Mani L. Bhaumik Institute for Theoretical Physics, Department of Physics and Astronomy,}}
\centerline{{\it University of California, Los Angeles, CA 90095, USA}}
\centerline{{\it ${}^c$ C.~N.~Yang Institute for Theoretical Physics,  Stony Brook University, Stony Brook,}}
\centerline{{\it NY 11794-3840, USA}}
\centerline{{\it ${}^d$ Simons Center for Geometry and Physics, Stony Brook University, Stony Brook,}}
\centerline{{\it NY 11794-3840, USA}}
\vspace{1cm}

\end{center}

\vskip 0.3 cm

\setcounter{footnote}{0}
\renewcommand{\thefootnote}{\arabic{footnote}}   
   
\begin{abstract}

Many interesting phenomena in quantum field theory such as dualities and symmetry enhancements can be understood using higher dimensional constructions. In this paper, we study compactifications of the rank $1$ $5d$ Seiberg $E_{N_f+1}$ SCFTs to $3d$ on Riemann surfaces of genus $g>1$. We rely on the recent progress in the study of compactifications of $6d$ SCFTs to $4d$ and torus compactifications of $5d$ SCFTs to conjecture $3d$ $\mathcal{N}=2$ theories corresponding to the reduction of said $5d$ SCFTs on three punctured spheres. These can then be used to build $3d$ $\mathcal{N}=2$ models corresponding to compactifications on more general surfaces. The conjectured theories are tested by comparing their properties against those expected from the compactification picture.

\end{abstract}
 
 \newpage
 
\tableofcontents

\section{Introduction}

Quantum field theory (QFT) is home to many mysterious phenomena. One example is the phenomenon of duality, where a single quantum field theory system can be represented in two different ways. There are several slightly different contexts where such a relation can manifest. One notable example is where the two dual descriptions emerge on different points of the same conformal manifold, that is the two descriptions differ by marginal deformations. Another notable example is where two different UV theories flow to the same IR theory. From the IR viewpoint then, the two describe different possible UV theories, that is they differ by irrelevant deformations.

Another example of peculiar QFT phenomena is enhancement of symmetry, where the global symmetry of the IR theory is bigger than that apparent from the underlying UV description. The additional symmetry is usually referred to as accidental symmetry, highlighting its unexpected nature. From the low energy viewpoint, it is manifested as a symmetry broken only by irrelevant deformations, and as such is expected to reemerge in the deep IR.

An interesting question is whether we can find some principle explaining such mysterious phenomena. One approach to tackling this is through the concept of dimensional reduction. The idea is to realize the theory through the compactification of a higher dimensional theory. An interesting feature in such a construction is that the resulting low energy theory generally depends only on certain properties of the surface, like its topology or complex structure, but is insensitive to other properties, usually geometric properties of the surface. The former properties then manifest as relevant or marginal deformations at low energies, while the latter appear as irrelevant ones.

This feature naturally makes dimensional reduction useful to understanding the phenomena described above. For instance, consider the dimensional reduction on two different surfaces. If the surfaces differ only through the properties that do not affect the IR theory, then the resulting low energy theories should differ only through irrelevant operators, that is they would be dual of the second type described above. If they, however, differ by properties associated with marginal operators, then we get a duality of the first type discussed. Similarly, we can consider compactification surfaces where a global symmetry is broken only by properties that do not affect the IR theory. This should then lead to a low energy theory where this symmetry is broken by irrelevant terms, that is this symmetry would appear accidental from the lower dimensional viewpoint, even though it is quite natural from the higher dimensional viewpoint.   

This property of dimensional reduction makes it extremely interesting to study and develop, and indeed it has been studied in a variety of different contexts. Out of the possible compactifications, reduction on 2d surfaces seems the most appealing. This is because they sit at the sweet spot where the geometry is sufficiently rich to allow multiple possibilities, while remaining rather straightforward and well-understood. The study of such compactifications was done mostly in the context of reductions of $6d$ SCFTs, for both $(2,0)$ \cite{Gaiotto:2009we,Gaiotto:2009hg,Benini:2009mz,Bah:2012dg} and $(1,0)$ supersymmetry \cite{Ganor:1996pc,Gaiotto:2015usa,Ohmori:2015pua,Ohmori:2015pia,Razamat:2016dpl,Bah:2017gph,Kim:2017toz,Kim:2018bpg,Kim:2018lfo,Razamat:2018gro,Razamat:2018gbu,Zafrir:2018hkr,Ohmori:2018ona,Sela:2019nqa,Chen:2019njf,Razamat:2019mdt,Pasquetti:2019hxf,Razamat:2019ukg,Razamat:2020bix,Sabag:2020elc,Baume:2021qho,Hwang:2021xyw,Distler:2022yse,Heckman:2022suy,Sabag:2022hyw} (see also \cite{Razamat:2022gpm} for a recent review). Despite the great progress on the study of compactifications in the $6d$ case, surface compactifications in other dimensions remains relatively unexplored. In the rest of this paper we shall be concerned with compactifications of $5d$ theories, specifically $5d$ SCFTs.\footnote{Besides $6d$ and $5d$, there are also some results on the surface compactifications of $4d$ \cite{Kutasov:2013ffl,Kutasov:2014hha,Putrov:2015jpa,Gadde:2015wta,Dedushenko:2017osi,Sacchi:2020pet} and $3d$ \cite{Benini:2022bwa} theories.}

Recently, progress was made also in understanding the compactification of $5d$ SCFTs to $3d$ $\mathcal{N}=2$ theories. Specifically, by relying on techniques that proved useful in tackling the reduction of $6d$ $(1,0)$ SCFTs, compactifications of $5d$ SCFTs on tori with flux have been relatively understood. This includes both the case of the Seiberg rank $1$ $E_{N_f+1}$ theories \cite{Sacchi:2021afk}, as well as their higher rank generalizations based on UV completions of $5d$ $SU$ type gauge theories \cite{Sacchi:2021wvg}. Nevertheless, so far only compactifications on tori were considered. The purpose of this paper is to extend this progress to compactifications on higher genus Riemann surfaces. The strategy we adopt is to again rely on the recent progress in understanding compactifications of $6d$ SCFTs to also understand similar compactifications of $5d$ SCFTs. 

Specifically, to understand compactifications on generic Riemann surfaces, we would need to understand theories associated with the compactification on three punctured spheres, the so-called trinion theories. From these, generic Riemann surfaces can be built by exploiting their pair-of-pants decomposition. Here, we shall consider trinion theories for $5d$ SCFTs resulting from the circle compactification of $6d$ SCFTs followed by mass deformations, for which trinion theories are already understood \cite{Razamat:2019ukg,Razamat:2020bix}. We can then hope that the compactification order commutes and opt to understand the trinions for the $5d$ SCFTs using the circle reduction of the $4d$ trinion theories of the associated $6d$ SCFTs. Alternatively, we can try to mimic the methods of \cite{Razamat:2019mdt,Razamat:2020bix}, used to determine the $4d$ trinions, and apply them to conjecture $3d$ trinions for the selected $5d$ SCFTs. As we shall show, both methods lead to the same set of $3d$ trinion theories. 

Once we have a guess for the $3d$ trinions, we can glue them together to build higher genus Riemann surfaces. If our conjecture is correct, that is the resulting theories describe the compactification of the $5d$ SCFTs on higher genus Riemann surfaces, then they should inherit various properties. These properties include the manifestation of the global symmetry expected from $5d$, as well as the presence of various operators descending from special multiplets in the parent $5d$ SCFT. This allows us to check said conjectural trinions by testing for the presence of these properties. In particular, as mentioned the $3d$ theory should inherit the global symmetry of the $5d$ SCFT. However, in many cases it is not manifest at the Lagrangian level, and only arises at the IR. In addition to providing consistency checks for this proposal,  these exhibit the application of the higher dimensional approach to the study of symmetry enhancement.  

The structure of this article is as follows. We begin in Section \ref{sec:review} with a short review on the compactifications of $5d$ SCFTs on tori and tubes. This will briefly introduce some known results regarding the compactifications of $5d$ SCFTs that will be used throughout this article. Section \ref{sec:trinion} deals with the derivation of the trinion theory for the Seiberg rank $1$ $E_{N_f+1}$ SCFTs. Once we acquire a conjecture for the trinion theory, we can try to test it by building $3d$ theories associated with the compactification on genus $g>1$ Riemann surfaces without punctures. This is carried out in Sections \ref{sec:noflux}, for cases without global symmetry fluxes, and \ref{sec:flux} for cases with ones. We end the paper with a short discussion.

\section{\boldmath Review of tubes and tori compactifications of $5d$ SCFTs}
\label{sec:review}

In this section we briefly review some of the results \cite{Sacchi:2021afk,Sacchi:2021wvg} on the compactification of certain $5d$ SCFTs on two punctured spheres, i.e.~tubes, and tori with flux. This serves two purposes: the first one is to introduce the reader to some of the concepts that are key in studying compactifications of $5d$ SCFTs to $3d$, and the second one is to review some results that we will then use in the next section to derive the trinion theory.

\subsection{Basic tube for the rank 1 $E_{N_f+1}$ SCFTs}
\label{subsec:tuberank1}

We start by considering the compactification of the rank 1 $E_{N_f+1}$ SCFTs on a tube with flux, which has been studied in \cite{Sacchi:2021afk}. The most fundamental of such tube compactifications, which we will call the \emph{basic tube} for brevity, is the one corresponding to a flux that in the $U(1)\times SO(2N_f)\subset E_{N_f+1}$ basis corresponds to\footnote{For more details regarding the basis and normalization of the fluxes used here, we refer the reader to Appendix A of \cite{Sacchi:2021afk}.}
\be\label{eq:fluxtube}
\mathcal{F}_{tube}=\Big(\frac{\sqrt{8-N_f}}{4};\underbrace{\frac{1}{4},\frac{1}{4},\cdots,\frac{1}{4}}_{N_f}\Big)\,.
\ee
The symmetry preserved by this flux is a subgroup of $E_{N_f+1}$ which for $0\leq N_f\leq 6$ is given by\footnote{The case of $N_f=7$ was not fully understood in \cite{Sacchi:2021afk} and so we neglect it here.}
\bea
N_f=6&:&\quad SO(12)\times U(1)\subset E_7\,,\nn\\
N_f=5&:&\quad SU(6)\times U(1)\subset E_6\,,\nn\\
N_f=4&:&\quad SU(4)\times SU(2)\times U(1)\subset E_5=SO(10)\,,\nn\\
N_f=3&:&\quad SU(3)\times U(1)^2\subset E_4=SU(5)\,,\nn\\
N_f=2&:&\quad SU(2)\times U(1)^2\subset E_3=SU(3)\times SU(2)\,,\nn\\
N_f=1&:&\quad U(1)^2\subset E_2=SU(2)\times U(1)\,,\nn\\
N_f=0&:&\quad U(1)\subset E_1=SU(2)\,.
\eea

\begin{figure}[t]
\center
\begin{tikzpicture}[baseline=0, font=\scriptsize]
\node[draw, rectangle] (l) at (0,0) {\large $\,\, 2\,\,$};
\node[draw, rectangle] (r) at (3,0) {\large $\,\, 2\,\,$};
\node[draw, rectangle] (b) at (1.5,-2.5) {\large $\, N_f\,$};
\draw[draw, solid,-] (l)--(r);

\draw[draw, solid,->] (l)--(0.75,-1.25);
\draw[draw, solid,-] (0.75,-1.25)--(b);
\draw[draw, solid,-<] (r)--(2.25,-1.25);
\draw[draw, solid,-] (2.25,-1.25)--(b);

\node[thick,rotate=0] at (1.5,0) {\Large$\times$};

\node[] at (1.5,0.5) {\normalsize $B,x^{\frac{2}{3}}a^{-2}$};
\node[] at (-0.2,-1.25) {\normalsize $Q,x^{\frac{2}{3}}ab$};
\node[] at (3.3,-1.25) {\normalsize $\tilde{Q},x^{\frac{2}{3}}ab^{-1}$};
\end{tikzpicture}
\caption{The $3d$ $\mathcal{N}=2$ Lagrangian of \cite{Sacchi:2021afk} for the compactification of the $5d$ rank 1 $E_{N_f+1}$ SCFT on a tube with flux $\mathcal{F}_{tube}$. Each square node denotes a flavor symmetry of special unitary type, while the charges under the abelian global symmetries including the R-symmetry are encoded in the powers of the corresponding fugacities. These are written close to each line that represents a chiral field transforming under the nodes they are connecting, with incoming arrows denoting the fundamental representation and outgoing the anti-fundamental. The cross denotes a single chiral multiplet $F$ with charges $x^{\frac{4}{3}}a^4$ which flips the meson constructed from the $SU(2)\times SU(2)$ bifundamental.}
\label{fig:3dtube}
\end{figure}
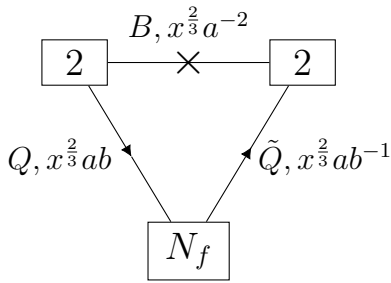 

The $3d$ $\mathcal{N}=2$ theory arising from this compactification is the Wess--Zumino (WZ) model depicted in Figure \ref{fig:3dtube}. The superpotential interaction between the chiral fields is
\be
\mathcal{W}=QB\tilde{Q}+FB^2\,,
\ee
where the contraction of flavor and gauge indices is understood. The symmetry manifest in this model is 
\be
SU(2)^2\times SU(N_f)\times U(1)_a\times U(1)_b\,.
\ee
The two $SU(2)$ symmetries are associated with each of the punctures, while the rest is a subgroup of the $5d$ $E_{N_f+1}$ symmetry. Notice that this is only a subgroup of the symmetry preserved by the flux due to the presence of the punctures. The full symmetry preserved by the flux is expected to get enhanced from $SU(N_f)\times U(1)_b$ when tubes are glued together so to form a Riemann surface with no punctures. The symmetry $U(1)_a$ is instead related to the $U(1)$ inside the $5d$ symmetry for which we turned on the flux.

The fields $Q$ and $\tilde{Q}$ are usually referred to as \emph{moment maps}. With this, one usually refers to operators that transform under the subgroup of the $5d$ symmetry, that is manifest in the $3d$ model, and under each of the $SU(2)$ puncture symmetries. These $3d$ operators are expected to descend from certain $5d$ operators that are given suitable boundary conditions at the location of the punctures, and are therefore associated with the punctures. This name is, as standard in the literature, given with an abuse of terminology, since in theories with four supercharges there is no precise notion of moment map operators. Notice that $Q$ and $\tilde{Q}$ transform in complex conjugate representations with respect to each other. This implies that the two punctures of the tube are of opposite type, usually referred to as \emph{sign} or \emph{color} in the literature.

\subsection{Gluing prescription}
\label{subsec:gluing}

Using the basic tube of Figure \ref{fig:3dtube} one can produce others with more general values of flux by chaining multiple tubes together to form one long tube. This adjoining of the tubes is done by gluing them along the punctures. In this subsection we summarize how such gluing is realized in field theory. 

First of all, when we glue two punctures we have to gauge the diagonal combination of their $SU(2)\times SU(2)$ symmetries. In the process we also need to turn on a Chern--Simons (CS) level in order to avoid the parity anomaly, which is taken to be at level $k=\pm\frac{6-N_f}{2}$ where the sign changes between adjacent gauge nodes in the final quiver.\footnote{The fact that a CS level is needed and the exact levels were realized by comparing against the expectations from the compactification picture. In particular, it was observed that for $N=6$ no CS term is needed, and smaller values of $N_f$ can then be realized by giving real masses to some of the flavors which generates the CS terms. As the origin of the CS term can be understood from the real mass deformations, in certain cases, notably when $S$-gluing to be defined below is involved, the CS level may differ.} 

\begin{figure}[t]
\center
\begin{tikzpicture}[baseline=0, font=\scriptsize]
\node[draw, rectangle] (l) at (0,0) {\large $\,\, 2\,\,$};
\node[draw, circle] (c) at (3,0) {\large $2_k$};
\node[draw, rectangle] (r) at (6,0) {\large $\,\, 2\,\,$};
\node[draw, rectangle] (b) at (3,-2.5) {\large $\, x\,$};
\node[draw, rectangle] (t) at (3,2.5) {\large $N_f-x$};
\draw[draw, solid,-] (l)--(c);
\draw[draw, solid,-] (c)--(r);

\draw[draw, solid,->] (l)--(1.5,-1.25);
\draw[draw, solid,-] (1.5,-1.25)--(b);
\draw[draw, solid,-<] (r)--(4.5,-1.25);
\draw[draw, solid,-] (4.5,-1.25)--(b);
\draw[draw, solid,->] (l)--(1.5,1.25);
\draw[draw, solid,-] (1.5,1.25)--(t);
\draw[draw, solid,-<] (r)--(4.5,1.25);
\draw[draw, solid,-] (4.5,1.25)--(t);
\draw[draw, solid,-<] (c)--(3,1.25);
\draw[draw, solid,-] (3,1.25)--(t);

\node[thick,rotate=0] at (1.5,0) {\Large$\times$};
\node[thick,rotate=0] at (4.5,0) {\Large$\times$};
\end{tikzpicture}
\caption{The $3d$ $\mathcal{N}=2$ Lagrangian of \cite{Sacchi:2021afk} for the compactification of the $5d$ rank 1 $E_{N_f+1}$ SCFT on a tube with flux $\mathcal{F}_{glue}$. We recall that each square node denotes a flavor symmetry, while the circle node now denotes a gauge symmetry, all of special unitary type. The $SU(2)$ gauge node also has a Chern--Simons level $k=\frac{6-N_f}{2}$. Note that here $x$ has to be even to be consistent with the parity anomaly.}
\label{fig:3dtubegen}
\end{figure}
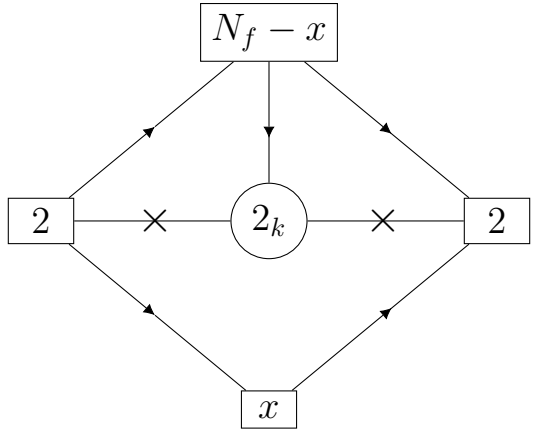 

We then need to turn on superpotential interactions that relate the moment maps of the glued punctures. For this we have two possible choices. The first one, called \emph{$\Phi$-gluing}, consists of adding $N_f$ fields $\Phi$ in the fundamental representation of the gauged $SU(2)$ symmetry with the interaction
\be
\gd\mathcal{W}=\Phi\cdot(Q+Q')=\sum_{a=1}^{N_f}\Phi^a(Q_a+Q'_a)\,,
\ee
where $Q$ and $Q'$ are the moment maps of the glued punctures of the two tubes and we are omitting the contraction of the gauge indices. When we perform a $\Phi$-gluing the fluxes of the component tubes add up, so that we get a flux which is twice the one \eqref{eq:fluxtube} of the basic tube.

The second possibility, called \emph{$S$-gluing}, consists of adding no extra chiral field and turning on the interaction
\be
\gd\mathcal{W}=Q\cdot Q'=\sum_{a=1}^{N_f}Q^aQ'_a\,.
\ee
In this case the fluxes of the component tubes get subtracted.

We can also consider mixed situations, in which we perform the $\Phi$-gluing for some of the moment maps and the $S$-gluing for the others
\be
\gd\mathcal{W}=\sum_{a=1}^{N_f-x}\Phi^a(Q_a+Q'_a)+\sum_{b=N_f-x+1}^{N_f}Q^bQ'_b\,,
\ee
where here $x$ has to be even. In this case we add the components of the flux vector corresponding to the moment maps for which we performed the $\Phi$-gluing and subtract those for which we performed the $S$-gluing. For instance, when gluing two copies of the basic tube in this way, the resulting flux is  
\be
\mathcal{F}_{glue}=\Big(\frac{\sqrt{8-N_f}}{2};\underbrace{\frac{1}{2},\cdots,\frac{1}{2}}_{N_f-x},\underbrace{0,\cdots,0}_{x}\Big)\,.
\ee
The resulting model is schematically depicted in Figure \ref{fig:3dtubegen} and the superpotential consists of cubic and quartic interactions corresponding to the three closed loops in the quiver.

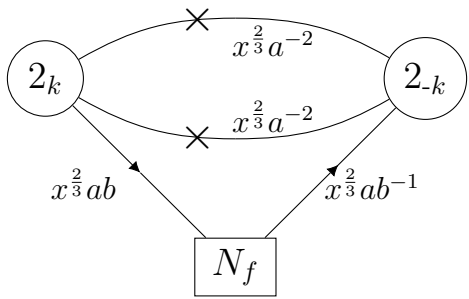
\begin{figure}[t]
\center
\begin{tikzpicture}[baseline=0, font=\scriptsize]
\node[draw, circle] (l) at (-1,0) {\large $ 2_k$};
\node[draw, circle] (r) at (4,0) {\large $ 2_{\text{-}k}$};
\node[draw, rectangle] (b) at (1.5,-2.5) {\large $\, N_f\,$};
\draw[draw, solid,-] (l) edge [out=30,in=180,loop,looseness=1] (1.5,0.8);
\draw[draw, solid,-] (l) edge [out=-30,in=180,loop,looseness=1] (1.5,-0.8);
\draw[draw, solid,-] (1.5,0.8) edge [out=0,in=150,loop,looseness=1] (r);
\draw[draw, solid,-] (1.5,-0.8) edge [out=0,in=-150,loop,looseness=1] (r);

\draw[draw, solid,->] (l)--(0.25,-1.25);
\draw[draw, solid,-] (0.25,-1.25)--(b);
\draw[draw, solid,-<] (r)--(2.75,-1.25);
\draw[draw, solid,-] (2.75,-1.25)--(b);

\node[thick,rotate=0] at (1,0.78) {\Large$\times$};
\node[thick,rotate=0] at (1,-0.78) {\Large$\times$};

\node[] at (-0.5,-1.4) {\normalsize $x^{\frac{2}{3}}ab$};
\node[] at (3.3,-1.4) {\normalsize $x^{\frac{2}{3}}ab^{-1}$};
\node[] at (2,0.5) {\normalsize $x^{\frac{2}{3}}a^{-2}$};
\node[] at (2,-0.5) {\normalsize $x^{\frac{2}{3}}a^{-2}$};
\end{tikzpicture}
\caption{The $3d$ $\mathcal{N}=2$ Lagrangian of \cite{Sacchi:2021afk} for the compactification of the $5d$ rank 1 $E_{N_f+1}$ SCFT on a torus. The Chern--Simons levels are written in terms of $k=\frac{6-N_f}{2}$.}
\label{fig:3dtorus}
\end{figure} 

With this gluing procedure we can also construct models corresponding to compactifications on tori. For example in Figure \ref{fig:3dtorus} we depict the theory corresponding to the compactification of the rank 1 $E_{N_f+1}$ SCFT on a torus with flux $\Big(\frac{\sqrt{8-N_f}}{2};\underbrace{\frac{1}{2},\frac{1}{2},\cdots,\frac{1}{2}}_{N_f}\Big)$, which is obtained by $\Phi$-gluing twice two copies of the basic tube. As pointed out in \cite{Sacchi:2021afk}, when constructing a surface with no punctures one typically has to turn on monopole superpotentials that break symmetries which would prevent the enhancement to the expected symmetry that is preserved by the flux. For the specific model of Figure \ref{fig:3dtorus} no monopole superpotential is actually needed.

\subsection{Tubes and tori for higher rank SCFTs}
\label{subsec:higherrank}

\begin{figure}[t]
\center
\begin{tikzpicture}[baseline=0, font=\scriptsize]
\node[draw, rectangle] (l) at (0,0) {\large $N+1$};
\node[draw, rectangle] (r) at (4,0) {\large $N+1$};
\node[draw, rectangle] (b) at (2,-2.5) {\large $\,\, m\,\,$};
\node[draw, rectangle] (t) at (2,2.5) {\large $N_f- m$};
\draw[draw, solid,-<] (l)--(2,0);
\draw[draw, solid](2,0)--(r);

\draw[draw, solid,->] (l)--(1,-1.25);
\draw[draw, solid,-] (1,-1.25)--(b);
\draw[draw, solid,-<] (r)--(3,-1.25);
\draw[draw, solid,-] (3,-1.25)--(b);
\draw[draw, solid,-<] (l)--(1,1.25);
\draw[draw, solid,-] (1,1.25)--(t);
\draw[draw, solid,->] (r)--(3,1.25);
\draw[draw, solid,-] (3,1.25)--(t);

\node[thick,rotate=0] at (1.2,0) {\Large$\times$};

\node[] at (2,0.5) {\normalsize $B$};
\node[] at (0.5,-1.25) {\normalsize $Q$};
\node[] at (3.5,-1.25) {\normalsize $\tilde{Q}$};
\node[] at (0.5,1.25) {\normalsize $q$};
\node[] at (3.5,1.25) {\normalsize $\tilde{q}$};
\end{tikzpicture}
\caption{The $3d$ $\mathcal{N}=2$ Lagrangian of \cite{Sacchi:2021wvg} for the compactification of the $5d$ rank $N$ SCFT that UV complete the $SU(N+1)_\kappa+N_fF$ gauge theories on a tube with flux.}
\label{fig:3dtuberankN}
\end{figure}
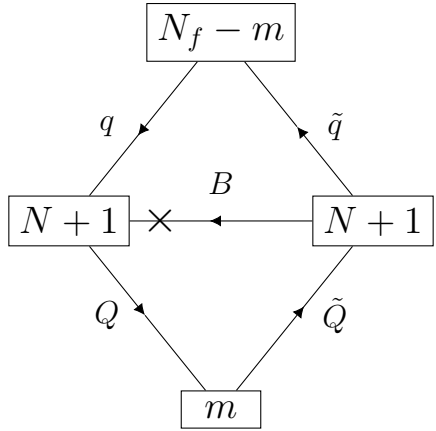 

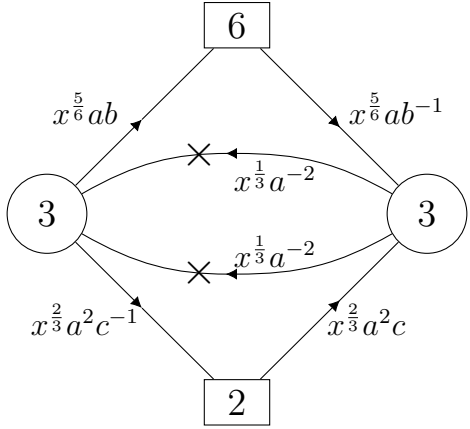
\begin{figure}[t]
\center
\begin{tikzpicture}[baseline=0, font=\scriptsize]
\node[draw, circle] (l) at (-1,0) {\large $\,\, 3\,\,$};
\node[draw, circle] (r) at (4,0) {\large $\,\, 3\,\,$};
\node[draw, rectangle] (b) at (1.5,-2.5) {\large $\,\, 2\,\,$};
\node[draw, rectangle] (t) at (1.5,2.5) {\large $\,\, 6\,\,$};
\draw[draw, solid,-<] (l) edge [out=30,in=180,loop,looseness=1] (1.5,0.8);
\draw[draw, solid,-<] (l) edge [out=-30,in=180,loop,looseness=1] (1.5,-0.8);
\draw[draw, solid,-] (1.5,0.8) edge [out=0,in=150,loop,looseness=1] (r);
\draw[draw, solid,-] (1.5,-0.8) edge [out=0,in=-150,loop,looseness=1] (r);

\draw[draw, solid,->] (l)--(0.25,-1.25);
\draw[draw, solid,-] (0.25,-1.25)--(b);
\draw[draw, solid,-<] (r)--(2.75,-1.25);
\draw[draw, solid,-] (2.75,-1.25)--(b);
\draw[draw, solid,->] (l)--(0.25,1.25);
\draw[draw, solid,-] (0.25,1.25)--(t);
\draw[draw, solid,-<] (r)--(2.75,1.25);
\draw[draw, solid,-] (2.75,1.25)--(t);

\node[thick,rotate=0] at (1,0.78) {\Large$\times$};
\node[thick,rotate=0] at (1,-0.78) {\Large$\times$};

\node[] at (-0.5,-1.4) {\normalsize $x^{\frac{2}{3}}a^2c^{-1}$};
\node[] at (3.2,-1.4) {\normalsize $x^{\frac{2}{3}}a^2c$};
\node[] at (-0.5,1.4) {\normalsize $x^{\frac{5}{6}}ab$};
\node[] at (3.6,1.4) {\normalsize $x^{\frac{5}{6}}ab^{-1}$};
\node[] at (2,0.5) {\normalsize $x^{\frac{1}{3}}a^{-2}$};
\node[] at (2,-0.5) {\normalsize $x^{\frac{1}{3}}a^{-2}$};
\end{tikzpicture}
\caption{The $3d$ $\mathcal{N}=2$ Lagrangian of \cite{Sacchi:2021wvg} for the compactification of the $5d$ rank 2 $SU(2)\times SO(16)$ SCFT on a torus with a unit of flux for a $U(1)$ inside $SO(16)$ whose commutant is $SO(12)\times SU(2)$.}
\label{fig:rank23dtube}
\end{figure}

Finally, let us quickly review some of the results of \cite{Sacchi:2021wvg} for the compactification of higher rank $5d$ SCFTs that UV complete the $SU(N+1)$ gauge theories with CS level $\kappa$ and $N_f$ fundamental hypermultiplets, which we compactly denote by $SU(N+1)_\kappa+N_fF$, for certain values of $\kappa$ and $N_f$. Some of the possible basic tubes in this case take the general form of Figure \ref{fig:3dtuberankN}. The superpotential is given by
\be
\mathcal{W}=QB\tilde{Q}+FB^2+qB^N\tilde{q}\,.
\ee
The value of $m$ generically depends both on $\kappa$ and on the value of the flux. For example in Figure \ref{fig:rank23dtube} we depict the model corresponding to the torus compactification of the $5d$ rank 2 SCFT with global symmetry $SU(2)\times SO(16)$ that UV completes the $SU(3)_1+8F$ gauge theory with the minimal allowed flux for a $U(1)$ inside $SO(16)$ whose commutant is $SO(12)\times SU(2)$. This model will play an important role in the derivation of the trinion theory that we will discuss in the next section. We refer the reader to \cite{Sacchi:2021wvg} for more details on the tube and tori compactifications of these higher rank $5d$ SCFTs.

\subsection{Tests}
\label{subsec:tests}

There are various tests that we can perform to check whether a particular $3d$ $\mathcal{N}=2$ Lagrangian theory flows to the same SCFT arising from the compactification of the $5d$ SCFT. These have been extensively discussed for example in Subsection 2.3 of \cite{Sacchi:2021afk} and here we shall only briefly review them. 

As we already mentioned, the $3d$ Lagrangians typically have a manifest symmetry which is smaller than the subgroup of the $5d$ symmetry that is preserved by the flux. One first possible test then consists in checking whether the full symmetry preserved by the flux gets enhanced at low energies in the $3d$ theory. This can be done in two ways. One consists of computing the supersymmetric index \cite{Bhattacharya:2008zy,Kim:2009wb,Imamura:2011su,Krattenthaler:2011da,Kapustin:2011jm,Willett:2016adv} and verifying that the spectrum of operators rearranges into characters of the expected enhanced symmetry. The second one consists in computing the flavor symmetry central charges, which can be extracted from the $S^3$ partition function \cite{Jafferis:2010un,Closset:2012vg,Gang:2019jut} (see Appendix B of \cite{Sacchi:2021afk} for a brief review), and verifying that their ratios are compatible with the symmetry enhancement, or more precisely with the ratios of the corresponding embedding indices inside the enhanced symmetry.

Another test consists of verifying that the spectrum of the $3d$ theory, which again can be investigated using the supersymmetric index, contains certain operators that descend from special $5d$ operators, such as the stress-energy tensor and the flavor symmetry current. If we compactify a $5d$ SCFT on a Riemann surface of genus $g$ with flux $F$ in a $U(1)_\alpha$ subgroup of its flavor symmetry $G$ that breaks it to $U(1)_{\alpha} \times \tilde{G}$ and with no punctures, then the index of the $3d$ theory computed with the reference $U(1)_R$ R-symmetry that is associated with the Cartan of the $5d$ $SU(2)_R$ R-symmetry is expected to take the following general form to low orders \cite{BRZtoapp}:
\bea \label{Indexexp}
I = & 1 & + (\sum_{i|q_i>0} \alpha^{q_i} \chi_{\mathcal{R}_i} (\tilde{G}) (g-1+q_i F))x^2 + \left(3g-3 + (1+\chi_{adj} (\tilde{G}))(g-1)\right)x^2 \nn\\ & + & (\sum_{i|q_i<0} \alpha^{q_i} \chi_{\mathcal{R}_i} (\tilde{G}) (g-1+q_i F))x^2 + \cdots\,,
\eea 
where $q_i$ and $\chi_{\mathcal{R}_i}$ are the charges and the characters of the representations under $U(1)_{\alpha} \times \tilde{G}$ that appear in the decomposition of the adjoint representation of $G$
\be \label{Gdecomp}
\chi_{adj} (G) = \sum_i \alpha^{q_i} \chi_{\mathcal{R}_i} (\tilde{G})\,.
\ee
Again we refer the reader to Subsubsection 2.3.2 of \cite{Sacchi:2021afk} for more explanations. One can also consider the contribution to the $3d$ index coming from other Higgs branch chiral generators rather than the flavor symmetry current. This was done in \cite{Sacchi:2021wvg}, see Subsection 2.3, though we will not need it in the present paper.

Let us also mention one last test that is possible in some cases, although we will not use it here. It can happen that different gluings of fundamental building blocks lead to distinct $3d$ theories associated with apparently different fluxes, but which are actually equivalent since they are related by an element of the Weyl group of the $5d$ flavor symmetry. Hence, the two compactifications should lead to the same $3d$ SCFT, meaning that the two different looking $3d$ Lagrangians are expected to be IR dual. Checking such a duality is another non-trivial test of the construction. For example in \cite{Sacchi:2021afk} some instances of this phenomenon were studied, in which the $3d$ IR duality turned out to be related to the Aharony duality \cite{Aharony:1997gp}.

\section{\boldmath Trinion for the compactification of the rank 1 $E_{N_f+1}$ SCFTs}
\label{sec:trinion}

In this section we derive $3d$ Lagrangian $\mathcal{N}=2$ theories that are conjectured to be the reduction of the rank 1 $E_{N_f+1}$ SCFTs on a three punctured sphere, which we shall refer to as \emph{trinions}. We shall consider two independent approaches, and see that they both give the same results for the trinion theory. 

\subsection{Conjecturing a Lagrangian for the trinion}

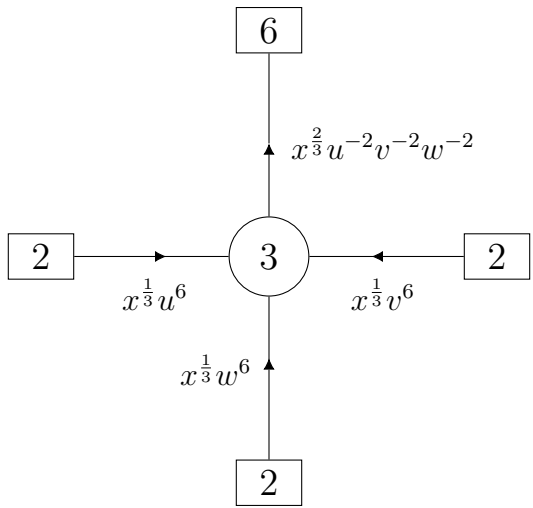
\begin{figure}[t]
\center
\begin{tikzpicture}[baseline=0, font=\scriptsize]
\node[draw, circle] (c) at (0,0) {\large $\,\, 3\,\,$};
\node[draw, rectangle] (t) at (0,3) {\large $\,\, 6\,\,$};
\node[draw, rectangle] (l) at (-3,0) {\large $\,\, 2\,\,$};
\node[draw, rectangle] (r) at (3,0) {\large $\,\, 2\,\,$};
\node[draw, rectangle] (b) at (0,-3) {\large $\,\, 2\,\,$};
\draw[draw, solid,->] (c)--(0,1.5);
\draw[draw, solid,-] (0,1.5)--(t);
\draw[draw, solid,-<] (c)--(-1.5,0);
\draw[draw, solid,-] (-1.5,0)--(l);
\draw[draw, solid,-<] (c)--(1.5,0);
\draw[draw, solid,-] (1.5,0)--(r);
\draw[draw, solid,-<] (c)--(0,-1.5);
\draw[draw, solid,-] (0,-1.5)--(b);

\node[] at (1.5,1.5) {\normalsize $x^{\frac{2}{3}}u^{-2}v^{-2}w^{-2}$};
\node[] at (-1.5,-0.5) {\normalsize $x^{\frac{1}{3}}u^6$};
\node[] at (1.5,-0.5) {\normalsize $x^{\frac{1}{3}}v^6$};
\node[] at (-0.7,-1.5) {\normalsize $x^{\frac{1}{3}}w^6$};

\end{tikzpicture}
\caption{The $4d$ $\mathcal{N}=1$ Lagrangian of \cite{Razamat:2020bix} for the compactification of the $6d$ rank 1 E-string SCFT on a three punctured sphere with unit flux for a $U(1)$ whose commutant in $E_8$ is $E_7$.}
\label{fig:4dtrinion}
\end{figure}

In order to conjecture a $3d$ Lagrangian for the compactification of the $5d$ rank 1 $E_{N_f+1}$ SCFTs on a three punctured sphere with flux we exploit the knowledge gained in similar compactifications, but in the $6d$ to $4d$ setup. Specifically, in \cite{Razamat:2020bix} it was proposed that the $4d$ $\mathcal{N}=1$ $SU(3)$ SQCD with 6 flavors is the theory obtained after compactifying the $6d$ rank 1 E-string theory on a three punctured sphere with unit flux for a $U(1)$ whose commutant inside the $6d$ $E_8$ global symmetry is $E_7$. In order to identify the embedding of the symmetries of the $4d$ model inside the $6d$ ones, it is useful to split the flavors as in Figure \ref{fig:4dtrinion}. The embedding is then
\bea
SU(6)\times U(1)^3\subset SU(6)\times SU(3)\times U(1)\subset E_7\times U(1)\subset E_8
\eea
and the flux is in the last $U(1)$. The remaining $SU(2)^3$ symmetries are associated with the three punctures.

We now want to consider the $5d$ SCFTs of rank 1 with $E_{N_f+1}$ global symmetry. One member of this family, corresponding to $N_f=7$ and $E_8$ global symmetry, can be obtained by compactifying the $6d$ E-string on a circle \cite{Ganor:1996pc}. From this $5d$ SCFT, the other members of the $E_{N_f+1}$ family can be produced by mass deformations, which from the $6d$ circle reduction viewpoint are interpreted as $E_{N_f+1}$ preserving holonomies in the $E_8$ global symmetry. 

This leads us to expect that we can obtain the $3d$ trinions we are looking for by taking the $4d$ trinion in Figure \ref{fig:4dtrinion}, compactifying it on a circle and turning on suitable real mass deformations in $3d$. The logic is as follows. We consider the reduction of the rank 1 E-string on $\Sigma\times S^1$, with an $E_{N_f+1}$ preserving holonomy around the $S^1$. Here, $\Sigma$ stands for some Riemann surface, potentially with flux in the $E_8$ global symmetry of the $6d$ SCFT. We can first reduce on the $S^1$ to get the $5d$ $E_{N_f+1}$ SCFT so the final theory should just be the compactification of this $5d$ SCFT on $\Sigma$. However, we can also consider reducing first on $\Sigma$ and then on the $S^1$. The first reduction now yields a $4d$ theory which is the compactification of the E-string SCFT on $\Sigma$. The second reduction then is just the compactification of this theory with the required holonomies, which become real mass deformations in $3d$. As long as the volumes of both surfaces are kept finite, the two orders of doing the reduction should yield the same result, and as we have been accustomed to the volume becoming an irrelevant deformation in dimensional reduction, it is reasonable to expect this to hold in this case. The following then is just the application of this idea to the case of $\Sigma$ being a three punctured sphere with the specified flux. 

From the above then, we expect the trinion theory for the rank 1 $E_8$ SCFT to just be the circle reduction of the $4d$ theory in Figure \ref{fig:4dtrinion}. The result of said reduction is a $3d$ $\mathcal{N}=2$ theory that looks exactly like the one in Figure \ref{fig:4dtrinion}, but with the fundamental $SU(3)$ monopole turned on in the superpotential \cite{Aharony:2013dha}. We shall actually not study this case in detail, due to the similarities with the $4d$ model, and rather concentrate on the cases with $N_f<7$. For these, we also need to incorporate real mass deformations.

It is convenient to start with the case of $N_f=6$. Here the appropriate real mass deformation turns out to be a positive real mass deformation to one of the six anti-fundamental chirals and a negative one to another. This can be understood by following the mapping of the symmetries between the $4d$ and $6d$ theories, worked out in \cite{Razamat:2020bix}, and seeing that this deformation correctly corresponds to the breaking of $E_8 \rightarrow E_7$. We shall also see that the resulting $3d$ theories indeed behave as expected from the reduction of the $5d$ $E_7$ SCFT. The resulting trinion theory is again a $3d$ $\mathcal{N}=2$ version of the one in Figure \ref{fig:4dtrinion}, but with only four anti-fundamental chirals and no monopole superpotential.

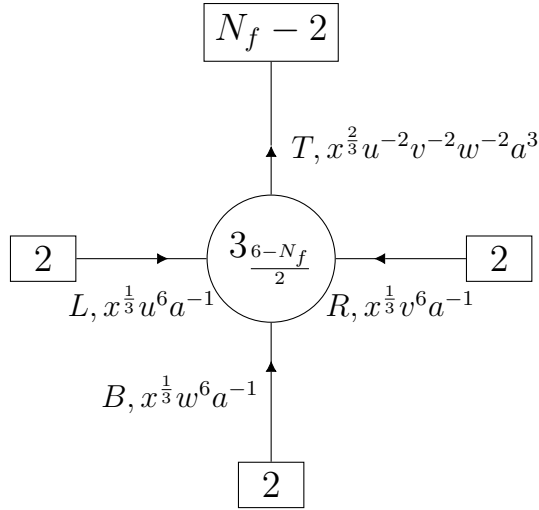
\begin{figure}[t]
\center
\begin{tikzpicture}[baseline=0, font=\scriptsize]
\node[draw, circle] (c) at (0,0) {\large $ 3_{\frac{6-N_f}{2}}$};
\node[draw, rectangle] (t) at (0,3) {\large $N_f-2$};
\node[draw, rectangle] (l) at (-3,0) {\large $\,\, 2\,\,$};
\node[draw, rectangle] (r) at (3,0) {\large $\,\, 2\,\,$};
\node[draw, rectangle] (b) at (0,-3) {\large $\,\, 2\,\,$};
\draw[draw, solid,->] (c)--(0,1.5);
\draw[draw, solid,-] (0,1.5)--(t);
\draw[draw, solid,-<] (c)--(-1.5,0);
\draw[draw, solid,-] (-1.5,0)--(l);
\draw[draw, solid,-<] (c)--(1.5,0);
\draw[draw, solid,-] (1.5,0)--(r);
\draw[draw, solid,-<] (c)--(0,-1.5);
\draw[draw, solid,-] (0,-1.5)--(b);

\node[] at (1.9,1.5) {\normalsize $T,x^{\frac{2}{3}}u^{-2}v^{-2}w^{-2}a^3$};
\node[] at (-1.7,-0.6) {\normalsize $L,x^{\frac{1}{3}}u^6a^{-1}$};
\node[] at (1.7,-0.6) {\normalsize $R,x^{\frac{1}{3}}v^6a^{-1}$};
\node[] at (-1.2,-1.8) {\normalsize $B,x^{\frac{1}{3}}w^6a^{-1}$};

\end{tikzpicture}
\caption{The $3d$ $\mathcal{N}=2$ Lagrangian for the compactification of the $5d$ rank 1 $E_{N_f+1}$ SCFT on a three punctured sphere with flux $(\frac{\sqrt{8-N_f}}{4};\frac{1}{4},\frac{1}{4},...,\frac{1}{4})$ in the $U(1)\times SO(2N_f)\subset E_{N_f+1}$ basis.}
\label{fig:3dtrinion}
\end{figure}

Next, we consider the cases with $N_f<6$. For this we need to turn on the real mass deformations implementing the required holonomies in the $6d$ construction. These turn out to be equivalent to integrating out $6-N_f$ anti-fundamentals with a positive mass, reducing the number of anti-fundamentals, and generating a CS term. The result is the theory in Figure \ref{fig:3dtrinion} for $2\leq N_f\leq 6$. Note that we cannot get to $N_f<2$ as we have only six anti-fundamentals, and integrating out the fundamentals would destroy the puncture symmetries. Nevertheless, we will see in the next subsection how we can still obtain theories with lower $N_f$ for a compactification on a genus two surface if there is no flux. 

We note that compared to the $4d$ trinion, we now have an additional abelian symmetry since the monopole superpotential has been lifted. This is consistent, since the rank of the manifest symmetry matches that of the $5d$ symmetry and the rank is preserved by the compactification. By analogy with $4d$, we expect the flux associated with this model in the $U(1)\times SO(2N_f)\subset E_{N_f+1}$ basis to be 
\be
\mathcal{F}_{tube}=(\frac{\sqrt{8-N_f}}{4};\underbrace{\frac{1}{4},\frac{1}{4},...,\frac{1}{4}}_{N_f})\,.
\ee
We also expect the $SU(2)^3$ symmetry to be associated with the three punctures, each carrying an $SU(2)$ global symmetry, while the $SU(N_f-2)\times U(1)^4$ global symmetry to be part of the $5d$ $E_{N_f+1}$ global symmetry. We shall see the exact embedding in the next section.

For each puncture, say the left one, we can also construct the moment maps as follows. We first have an operator in the fundamental representation of $SU(N_f-2)$ and of $SU(2)$, which is constructed as the meson $TL$ and similarly for the other punctures. We then have two operators which are singlets under $SU(N_f-2)$ and are in the fundamental of $SU(2)$, which are constructed as the mesons $R^2L$ and $B^2L$. These fields will be relevant when gluing the punctures of the trinion, so for convenience we will collectively denote them as follows:
\bea\label{eq:momentmaps}
&&M_L=\left(TL,R^2L,B^2L\right)\,,\nn\\
&&M_R=\left(TR,L^2R,B^2R\right)\,,\nn\\
&&M_B=\left(TB,R^2B,L^2B\right)\,.
\eea

In the next sections we will start gluing various copies of these theories so to construct surfaces of higher genus and without punctures. This serves for two purposes: on the one hand it allows us to perform tests for the conjecture that the one in Figure \ref{fig:3dtrinion} is the correct trinion model with the claimed flux, and on the other hand from our analysis we will understand what monopole superpotential is needed in the gluing. That a monopole superpotential is needed was indeed understood already in tube and tori compactifications in \cite{Sacchi:2021afk,Sacchi:2021wvg}. In the reminder of this section we present an additional derivation of the trinion model of Figure \ref{fig:3dtrinion}.

\subsection{From the tube to the trinion}

It is possible to derive the trinion model from a tube model by turning on a deformation triggered by a vev for a suitable operator. This approach was first discussed in the context of compactifications from $6d$ to $4d$ in \cite{Razamat:2019mdt} (see also \cite{Razamat:2019ukg,Sabag:2020elc} and Section 6.4 of \cite{Razamat:2022gpm} for a review). The idea is based on the fact that SCFTs (either in $6d$ or $5d$) of different ranks can be related to each other by turning on vevs for suitable Higgs branch operators. For example, in the case of the $6d$ minimal $(D_{N+3},D_{N+3})$ conformal matter one can flow to the one with rank decreased by one unit $(D_{N+2},D_{N+2})$ by turning on a vev for a Higgs branch operator which, in the gauge theory phase on the tensor branch in terms of a $USp(2N-2)$ gauge theory with $2N+6$ fundamental hypermultiplets, is described by one component of the meson matrix. Let us call $\mathcal{T}_{UV}^{6d/5d}$ and $\mathcal{T}_{IR}^{6d/5d}$ respectively the theories before and after the deformation. One might then wonder whether this deformation and the compactification process commute. Specifically, one can identify an operator in the theory $\mathcal{T}_{UV}^{4d/3d}$ (either in $4d$ or $3d$) obtained from the compactification of $\mathcal{T}_{UV}^{6d/5d}$ that descends from the aformentioned Higgs branch operator, deform the theory by a vev for it and ask whether the resulting theory $\mathcal{T}_{IR}^{4d/3d}$ can be interpreted as a compactification of the theory $\mathcal{T}_{IR}^{6d/5d}$. It turns out that the answer is yes, but the Riemann surface for the compactification $\mathcal{T}_{IR}^{6d/5d}\to \mathcal{T}_{IR}^{4d/3d}$ has an additional minimal puncture compared to the one for the compactification $\mathcal{T}_{UV}^{6d/5d}\to \mathcal{T}_{UV}^{4d/3d}$.

We can apply this logic to try to derive the trinion for the $5d$ rank 1 $E_{N_f+1}$ SCFT, which is the UV completion of the $SU(2)+N_fF$ gauge theory and which can be obtained from circle reduction plus decoupling of the $6d$ E-string theory that also corresponds to the $(D_4,D_4)$ conformal matter. For this we can start from the tube for some $5d$ rank 2 SCFT that can be obtained from circle reduction plus decoupling of the $6d$ $(D_5,D_5)$ conformal matter, which was discussed in \cite{Sacchi:2021wvg}. For simplicity we will consider the case of $N_f=6$, since the trinions for lower $N_f$ can be obtained from real mass deformations in $3d$. In this case, the $5d$ SCFT we will consider is the UV completion of the $SU(3)_1+8F$ gauge theory and in this description the vev leading to the $SU(2)+6F$ theory that we want to consider is for one component of the meson matrix. Moreover, we will actually study the vev deformation in a torus model, which should lead to the compactification of the lower rank theory on a torus with one extra puncture. We will then cut the torus open to obtain the trinion.

Our starting point is then the compactification of the $5d$ rank 2 SCFT with global symmetry $SU(2)\times SO(16)$, which is the UV completion of the $SU(3)_1+8F$ gauge theory, on a torus. A $3d$ quiver description for this model was proposed in \cite{Sacchi:2021wvg} and we already reported it in Figure \ref{fig:rank23dtube}. The charge assignment is due to two types of superpotential terms, one is a cubic interaction involving the upper triangle and one is a quartic interaction involving the lower triangle. The crosses denote singlet fields which flip the baryonic operators constructed from the associated bifundamental fields, though these won't play a role here. The reason is that in general, the low-energy theory when going on the moduli space includes singlets in addition to a potential interacting SCFT. As such, it is quite reasonable that some singlets in $\mathcal{T}_{IR}^{4d/3d}$ would have their origin from said singlets generated in the $6d$ flow and so should be removed. In $4d$ the identification of these singlets can be done by matching anomalies. However, we do not possess this tool in $3d$ and so have to rely on the $4d$ results as well as the consistency conditions in $3d$. In particular, these singlets are indeed removed in $4d$, and we shall see that removing them yields consistent results with the trinion of Figure \ref{fig:3dtrinion}.

We deform this theory by turning on a vev for one component of the mesonic operator constructed from the fields represented by the top right and the bottom right diagonal edges of the quiver, which has the effect of partially breaking the global symmetry and Higgsing the right $SU(3)$ gauge node down to $SU(2)$. The latter will then be interpreted as the diagonal gauging of the $SU(2)$ symmetries of the punctures of the trinion that are glued to form the torus. 

The effect of the vev can be effectively studied using supersymmetric partition functions, such as the $S^3_b$ partition function or the supersymmetric index, following the same strategy discussed in \cite{Gaiotto:2012xa} for the $4d$ supersymmetric index. At the level of the index, the deformation implies the constraint on the fugacities
\be\label{eq:VEVconstr}
a^3b^{-1}cx^{\frac{3}{2}}f_1^{-1}v=1\,,
\ee
where $f_i$ with $\prod_{i=1}^6f_i=1$ are the fugacities for the $SU(6)$ symmetry while $v$ is the fugacity for the $SU(2)$ symmetry. Such a constraint is obtained by requiring that the operator that gets the vev is uncharged under all the symmetries including the R-symmetry and it encodes the aforementioned breaking of the global symmetry. Imposing this constraint on the index one can see that the sets of poles coming from the contributions of the two chirals forming the meson to which we gave the vev pinch the integration contour at one point. We should then take the residue at such point for one of the integration variables of the right $SU(3)$ gauge node, which implements the Higgsing at the level of the index. 

\begin{figure}[t]
\center
\begin{tikzpicture}[baseline=0, font=\scriptsize]
\node[draw, circle] (l) at (-1,0) {\large $\,\, 3\,\,$};
\node[draw, circle] (r) at (4,0) {\large $\,\, 2\,\,$};
\node[draw, rectangle] (b1) at (-2.5,-2.5) {\large $\,\, 1\,\,$};
\node[draw, rectangle] (b2) at (-1,-2.5) {\large $\,\, 1\,\,$};
\node[draw, rectangle] (b3) at (0.5,-2.5) {\large $\,\, 1\,\,$};
\node[draw, rectangle] (b4) at (4,-2.5) {\large $\,\, 1\,\,$};
\node[draw, rectangle] (t) at (1.5,2.5) {\large $\,\, 5\,\,$};
\draw[draw, solid,-<] (l) edge [out=30,in=180,loop,looseness=1] (1.5,0.8);
\draw[draw, solid,-<] (l) edge [out=-30,in=180,loop,looseness=1] (1.5,-0.8);
\draw[draw, solid,-] (1.5,0.8) edge [out=0,in=150,loop,looseness=1] (r);
\draw[draw, solid,-] (1.5,-0.8) edge [out=0,in=-150,loop,looseness=1] (r);

\draw[draw, solid,-<] (l)--(-1.75,-1.25);
\draw[draw, solid,-] (-1.75,-1.25)--(b1);
\draw[draw, solid,-<] (l)--(-1,-1.25);
\draw[draw, solid,-] (-1,-1.25)--(b2);
\draw[draw, solid,-<] (l)--(-0.25,-1.25);
\draw[draw, solid,-] (-0.25,-1.25)--(b3);
\draw[draw, solid,-] (r)--(4,-1.25);
\draw[draw, solid,-] (4,-1.25)--(b4);
\draw[draw, solid,->] (l)--(0.25,1.25);
\draw[draw, solid,-] (0.25,1.25)--(t);
\draw[draw, solid,-<] (r)--(2.75,1.25);
\draw[draw, solid,-] (2.75,1.25)--(t);

\node[] at (-1,-3.1) {\small $x^{\frac{13}{6}}a^5b^{-1}s^{-5}$};
\node[] at (-3.5,-3.1) {\small $x^{\frac{7}{6}}a^{-1}b^{-1}s^{-5}$};
\node[] at (1.5,-3.1) {\small $x^{-\frac{5}{6}}a^{-1}bc^{-2}s^{5}$};
\node[] at (2.7,-1.4) {\small $x^{\frac{7}{4}}a^{\frac{9}{2}}b^{-\frac{1}{2}}c^2s^{-\frac{5}{2}}$};
\node[] at (-0.5,1.4) {\small $x^{\frac{5}{6}}abs^{-1}$};
\node[] at (4,1.4) {\small $x^{\frac{5}{4}}a^{\frac{3}{2}}b^{-\frac{3}{2}}s^{-\frac{3}{2}}$};
\node[] at (1.8,0.35) {\small $x^{-\frac{1}{12}}a^{-\frac{5}{2}}b^{\frac{1}{2}}s^{\frac{5}{2}}$};
\node[] at (1.8,-0.35) {\small $x^{-\frac{1}{12}}a^{-\frac{5}{2}}b^{\frac{1}{2}}s^{\frac{5}{2}}$};
\end{tikzpicture}
\begin{tikzpicture}[baseline=0, font=\scriptsize]
\node[draw, circle] (l) at (-1,0) {\large $\,\, 3\,\,$};
\node[draw, circle] (r) at (4,0) {\large $\,\, 2\,\,$};
\node[draw, rectangle] (b1) at (-2,-2.5) {\large $\,\, 2\,\,$};
\node[draw, rectangle] (b3) at (0,-2.5) {\large $\,\, 1\,\,$};
\node[draw, rectangle] (b4) at (4,-2.5) {\large $\,\, 1\,\,$};
\node[draw, rectangle] (t) at (1.5,2.5) {\large $\,\, 5\,\,$};
\draw[draw, solid,-<] (l) edge [out=30,in=180,loop,looseness=1] (1.5,0.8);
\draw[draw, solid,-<] (l) edge [out=-30,in=180,loop,looseness=1] (1.5,-0.8);
\draw[draw, solid,-] (1.5,0.8) edge [out=0,in=150,loop,looseness=1] (r);
\draw[draw, solid,-] (1.5,-0.8) edge [out=0,in=-150,loop,looseness=1] (r);

\draw[draw, solid,-<] (l)--(-1.5,-1.25);
\draw[draw, solid,-] (-1.5,-1.25)--(b1);
\draw[draw, solid,-<] (l)--(-0.5,-1.25);
\draw[draw, solid,-] (-0.5,-1.25)--(b3);
\draw[draw, solid,-] (r)--(4,-1.25);
\draw[draw, solid,-] (4,-1.25)--(b4);
\draw[draw, solid,->] (l)--(0.25,1.25);
\draw[draw, solid,-] (0.25,1.25)--(t);
\draw[draw, solid,-<] (r)--(2.75,1.25);
\draw[draw, solid,-] (2.75,1.25)--(t);

\node[] at (-2,-3.1) {\small $x^{\frac{1}{6}}a^{-1}c^{-1}$};
\node[] at (0,-3.1) {\small $x^{\frac{7}{6}}a^{5}c^{-1}$};
\node[] at (3.2,-1.4) {\small $x^{\frac{5}{4}}a^{\frac{9}{2}}c^{\frac{3}{2}}$};
\node[] at (-0.7,1.4) {\small $x^{\frac{19}{30}}ab^{\frac{6}{5}}c^{-\frac{1}{5}}$};
\node[] at (4.2,1.4) {\small $x^{\frac{19}{20}}a^{\frac{3}{2}}b^{-\frac{6}{5}}c^{-\frac{3}{10}}$};
\node[] at (1.8,0.35) {\small $x^{\frac{5}{12}}a^{-\frac{5}{2}}c^{\frac{1}{2}}$};
\node[] at (1.8,-0.35) {\small $x^{\frac{5}{12}}a^{-\frac{5}{2}}c^{\frac{1}{2}}$};

\node[] at (-2.2,-1.5) {\small $Q$};
\node[] at (0.2,-1.5) {\small $P$};
\node[] at (4.3,-1.4) {\small $T$};
\node[] at (-0.2,2) {\small $L$};
\node[] at (3.7,2) {\small $R$};
\node[] at (1.8,1) {\small $B_1$};
\node[] at (1.8,-1.05) {\small $B_2$};
\end{tikzpicture}
\caption{The result of the Higgsing induced by the vev in two different parametrizations of the global symmetries. In the bottom figure we also report the names we will use to denote each chiral fields.}
\label{fig:afterVEV}
\end{figure}
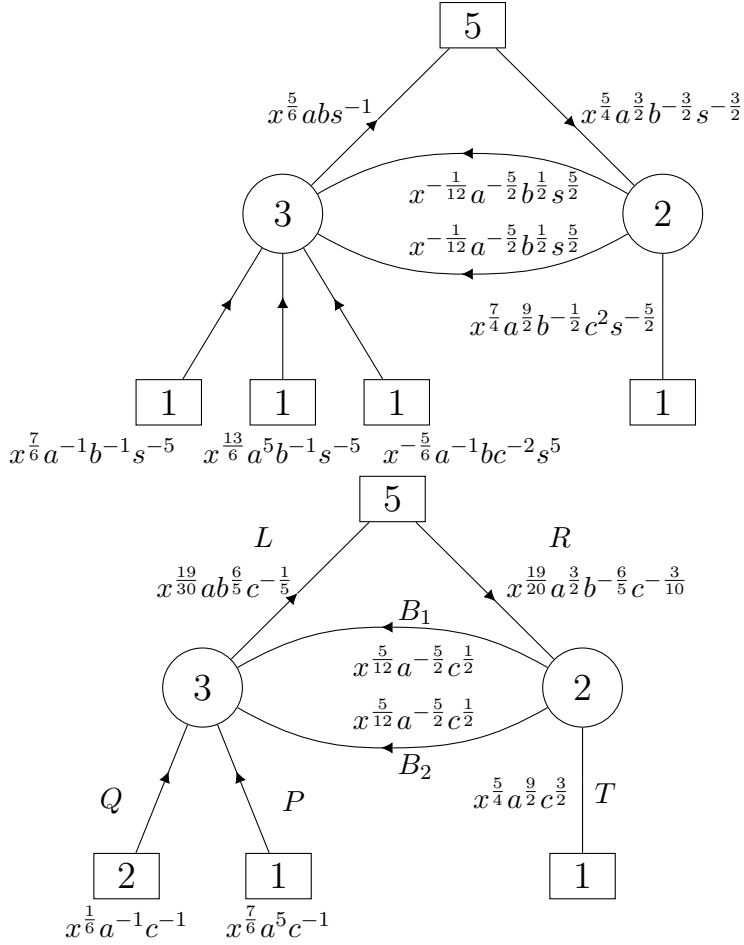

We choose to solve the constraint \eqref{eq:VEVconstr} by
\be
v=x^{-\frac{3}{2}}a^{-3}bc^{-1}f_1
\ee
and to take the residue 
\be
z^{(R)}_1=x^{-\frac{5}{6}}a^{-1}bf_1\,,
\ee
where $z^{(R)}_a$ with $\prod_{a=1}^3z^{(R)}_a=1$ are the gauge fugacities for the right $SU(3)$ node. Moreover, since the deformation breaks $SU(6)\to SU(5)\times U(1)$, we redefine for convenience
\be
f_i=\begin{cases}s^5 & i=1 \\ s^{-1}f_{i-1} & i=2,\cdots,6\end{cases}\,.
\ee
so that now $f_i$ with $\prod_{i=1}^5f_i=1$ are the $SU(5)$ fugacities. The result is the quiver on the top of Figure \ref{fig:afterVEV}.

As we previously mentioned, we expect this to be the compactification of the $E_7$ SCFT on a torus with one puncture. There should then be an $SU(2)$ symmetry associated with this new puncture that emerged at the end of the flow triggered by the VEV. This can be made manifest by performing the redefinion
\be 
s\to s^{\frac{1}{5}}x^{-\frac{1}{5}}b^{-\frac{1}{5}}c^{\frac{1}{5}},\quad a\to as^{\frac{1}{6}},\quad c\to cs^{-\frac{1}{6}}\,,
\ee
where now $s$ is the fugacity for such $SU(2)$ symmetry. The result is depicted on the bottom of Figure \ref{fig:afterVEV}.

Let us briefly comment on the superpotential of this model. This is given by
\be
\mathcal{W}=LR(B_1+B_2)+P(B_1^2+B_2^2)+TQ^2(B_1+B_2)\,,
\ee
where the names we use for the chiral fields are as specified in the figure. In particular, there is no monopole superpotential. Moreover, the fields $R$ and $T$ attached to the $SU(2)$ gauge node can be understood as the fields added in the $\Phi$-gluing of the trinion to form the torus with one puncture and which flip the moment map operators of the glued punctures, which are given by
\be
M_{1/2}=\left(LB_{1/2},Q^2B_{1/2}\right)\,.
\ee

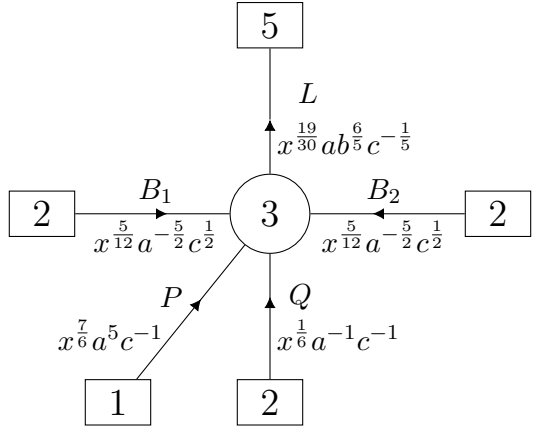
\begin{figure}[t]
\center
\begin{tikzpicture}[baseline=0, font=\scriptsize]
\node[draw, circle] (c) at (-1,0) {\large $\,\, 3\,\,$};
\node[draw, rectangle] (r) at (2,0) {\large $\,\, 2\,\,$};
\node[draw, rectangle] (l) at (-4,0) {\large $\,\, 2\,\,$};
\node[draw, rectangle] (b1) at (-1,-2.5) {\large $\,\, 2\,\,$};
\node[draw, rectangle] (b2) at (-3,-2.5) {\large $\,\, 1\,\,$};
\node[draw, rectangle] (t) at (-1,2.5) {\large $\,\, 5\,\,$};

\draw[draw, solid,-<] (c)--(-2.5,0);
\draw[draw, solid,-] (-2.5,0)--(l);
\draw[draw, solid,-<] (c)--(0.5,0);
\draw[draw, solid,-] (0.5,0)--(r);
\draw[draw, solid,-<] (c)--(-1,-1.25);
\draw[draw, solid,-] (-1,-1.25)--(b1);
\draw[draw, solid,-<] (c)--(-2,-1.25);
\draw[draw, solid,-] (-2,-1.25)--(b2);
\draw[draw, solid,->] (c)--(-1,1.25);
\draw[draw, solid,-] (-1,1.25)--(t);

\node[] at (-0.1,-1.6) {\small $x^{\frac{1}{6}}a^{-1}c^{-1}$};
\node[] at (-3.1,-1.6) {\small $x^{\frac{7}{6}}a^{5}c^{-1}$};
\node[] at (0,1) {\small $x^{\frac{19}{30}}ab^{\frac{6}{5}}c^{-\frac{1}{5}}$};
\node[] at (-2.5,-0.3) {\small $x^{\frac{5}{12}}a^{-\frac{5}{2}}c^{\frac{1}{2}}$};
\node[] at (0.5,-0.3) {\small $x^{\frac{5}{12}}a^{-\frac{5}{2}}c^{\frac{1}{2}}$};

\node[] at (-0.6,-1.1) {\small $Q$};
\node[] at (-2.3,-1.1) {\small $P$};
\node[] at (-0.5,1.6) {\small $L$};
\node[] at (-2.5,0.3) {\small $B_1$};
\node[] at (0.5,0.3) {\small $B_2$};
\end{tikzpicture}
\caption{The trinion model obtained by opening the torus with one puncture of Figure \ref{fig:afterVEV}.}
\label{fig:trinionVEV}
\end{figure}

\begin{figure}[b]
\center
\begin{tikzpicture}[baseline=0, font=\scriptsize]
\node[draw, circle] (c) at (-1,0) {\large $\,\, 3\,\,$};
\node[draw, circle] (r) at (2,0) {\large $\,\, 2\,\,$};
\node[draw, rectangle] (l) at (-4,0) {\large $\,\, 2\,\,$};
\node[draw, rectangle] (b1) at (-1,-2.5) {\large $\,\, 2\,\,$};
\node[draw, rectangle] (b2) at (-3,-2.5) {\large $\,\, 1\,\,$};
\node[draw, rectangle] (t) at (-1,2.5) {\large $\,\, 4\,\,$};
\node[draw, rectangle] (t2) at (2,2.5) {\large $\,\, 1\,\,$};
\node[draw, rectangle] (r2) at (5,0) {\large $\,\, 2\,\,$};

\draw[draw, solid,-<] (c)--(-2.5,0);
\draw[draw, solid,-] (-2.5,0)--(l);
\draw[draw, solid,-<] (c)--(0.5,0);
\draw[draw, solid,-] (0.5,0)--(r);
\draw[draw, solid,-<] (c)--(-1,-1.25);
\draw[draw, solid,-] (-1,-1.25)--(b1);
\draw[draw, solid,-<] (c)--(-2,-1.25);
\draw[draw, solid,-] (-2,-1.25)--(b2);
\draw[draw, solid,->] (c)--(0.5,1.25);
\draw[draw, solid,-] (0.5,1.25)--(t2);
\draw[draw, solid,->] (c)--(-1,1.25);
\draw[draw, solid,-] (-1,1.25)--(t);
\draw[draw, solid,-] (r)--(r2);
\draw[draw, solid,->] (r2)--(3.5,1.25);
\draw[draw, solid,-] (3.5,1.25)--(t2);
\draw[draw, solid,-<] (r)--(2,1.25);
\draw[draw, solid,-] (2,1.25)--(t2);

\node[] at (-0.1,-1.6) {\small $x^{\frac{1}{6}}a^{-1}c^{-1}$};
\node[] at (-3.1,-1.6) {\small $x^{\frac{7}{6}}a^{5}c^{-1}$};
\node[] at (-2,1.7) {\small $x^{\frac{19}{30}}ab^{\frac{6}{5}}c^{-\frac{1}{5}}$};
\node[] at (0,1.7) {\small $x^{\frac{19}{30}}ab^{\frac{6}{5}}c^{-\frac{1}{5}}$};
\node[] at (-2.5,-0.3) {\small $x^{\frac{5}{12}}a^{-\frac{5}{2}}c^{\frac{1}{2}}$};
\node[] at (0.5,-0.3) {\small $x^{\frac{5}{12}}a^{-\frac{5}{2}}c^{\frac{1}{2}}$};
\node[] at (3.5,-0.3) {\small $x^{-\frac{1}{10}}a^{3}b^{\frac{3}{5}}c^{-\frac{3}{5}}$};
\node[] at (1,0.7) {\footnotesize $x^{\frac{19}{20}}a^{\frac{3}{2}}b^{-\frac{6}{5}}c^{\frac{3}{10}}$};
\node[] at (4.6,1.7) {\small $x^{\frac{23}{20}}a^{-\frac{9}{2}}b^{\frac{3}{5}}c^{\frac{9}{10}}$};

\end{tikzpicture}
\caption{The gluing of the trinion of Figure \ref{fig:trinionVEV} and of the tube of Figure \ref{fig:3dtube}. Notice that the $SU(2)$ gauge nodes sees 6 chirals and has a monopole superpotential, so it confines giving the trinion of Figure \ref{fig:3dtrinion}.}
\label{fig:gluetubetrinion}
\end{figure}
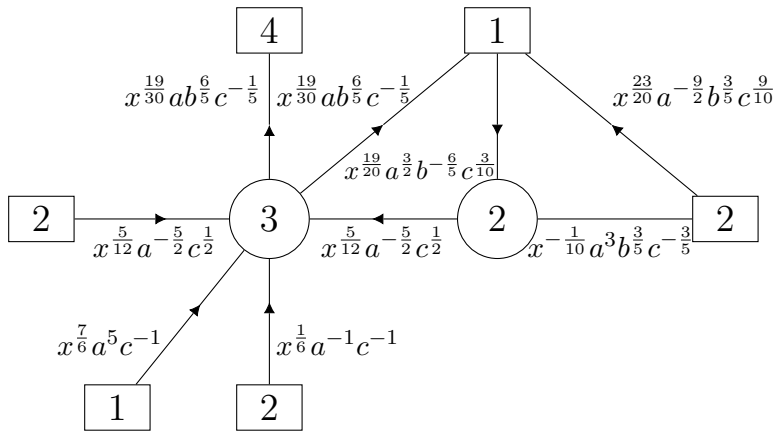

We now want to get the trinion by cutting the torus open. This is done by ungauging the $SU(2)$ symmetries and removing the fields $R$ and $T$ that are introduced in the $\Phi$-gluing. The resulting model is depicted in Figure \ref{fig:trinionVEV} and the superpotential is given by
\be\label{eq:superpottrinionVEV}
\mathcal{W}=P(B_1^2+B_2^2)\,.
\ee
Notice that this is not exactly the trinion of Figure \ref{fig:3dtrinion} that we discussed in the previous subsection. In order to get the desired trinion, we need to glue to one of the punctures of that in Figure \ref{fig:trinionVEV} one copy of the tube theory discussed in \cite{Sacchi:2021afk} and which we depicted in Figure \ref{fig:3dtube}, in a similar manner to the case of the $6d$ to $4d$ compactification of the $(D_{N+3},D_{N+3})$ conformal matter reviewed in Section 6.4 of \cite{Razamat:2022gpm}. More precisely we first flip the puncture of the trinion that we want to glue, that is we introduce singlet fields that flip the moment map operators $LB_{2}$ and $Q^2B_{2}$, and then we glue the tube with $\Phi$-gluing for all the moment maps except one for which we perform an $S$-gluing. The result is depicted in Figure \ref{fig:gluetubetrinion}. In the gluing we also turn on the fundamental monopole of the $SU(2)$ gauge node so that, using the duality of \cite{Aharony:2013dha,Aharony:2013kma},\footnote{More precisely, the $SU(2)$ gauge theory with 6 fundamental chirals and $\mathcal{W}=\mathfrak{M}$ is dual to a WZ model of 15 chirals with a cubic superpotential.} it confines. One of the fields produced in the dualization is mapped to the baryonic operator $(B_2)^2$ of the theory before the dualization and which becomes massive together with the field $P$ due to the superpotential \eqref{eq:superpottrinionVEV}. Integrating out all the massive fields one gets exactly the expected trinion of Figure \ref{fig:3dtrinion} (up to a reparametrization of the abelian symmetries and the R-symmetry).

\section{Compactifications on higher genus surfaces with no flux}
\label{sec:noflux}

After establishing the trinion model, we can now turn to study the theories resulting from the compactification of the $5d$ $E_{N_f+1}$ SCFTs on a higher genus surface constructed by gluing trinions. In this section we consider the case in which the punctures of the trinions are glued via $S$-gluing, as reviewed in Subsection \ref{subsec:gluing}, resulting in surfaces with no flux. We will begin by considering the case of genus 2 and then turn to the case of arbitrary higher genus.

\subsection{Genus two}

We first construct the models for a compactification on a genus 2 surface with no flux by gluing two copies of the trinion we introduced in Figure \ref{fig:3dtrinion}. We will start by considering the cases with $N_f\geq 2$ and then flow to lower $N_f$ by studying suitable mass deformations.

\subsubsection{$N_f\geq 2$}

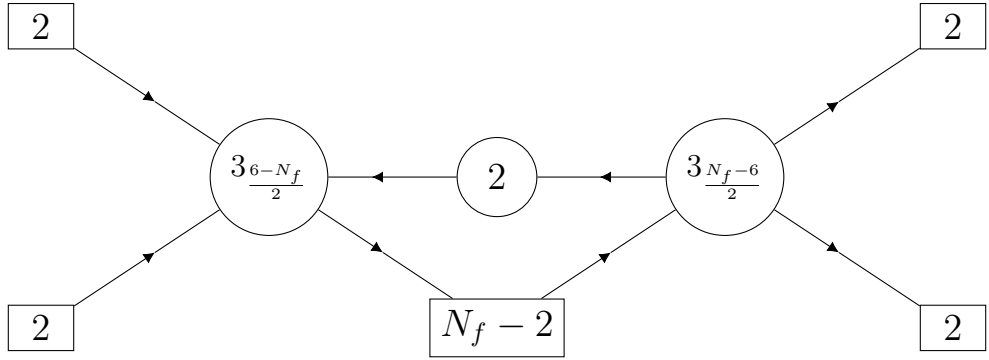
\begin{figure}[t]
\center
\begin{tikzpicture}[baseline=0, font=\scriptsize]
\node[draw, rectangle] (p1) at (0,2) {\large $\,\, 2\,\,$};
\node[draw, rectangle] (p2) at (0,-2) {\large $\,\, 2\,\,$};
\node[draw, circle] (p3) at (3,0) {\normalsize $3_{\frac{6-N_f}{2}}$};
\node[draw, circle] (p4) at (6,0) {\large $\,\, 2\,\,$};
\node[draw, rectangle] (p5) at (6,-2) {\large $N_f-2$};
\node[draw, circle] (p6) at (9,0) {\normalsize $3_{\frac{N_f-6}{2}}$};
\node[draw, rectangle] (p7) at (12,2) {\large $\,\, 2\,\,$};
\node[draw, rectangle] (p8) at (12,-2) {\large $\,\, 2\,\,$};

\draw[draw, solid,->] (p1)--(1.5,1);
\draw[draw, solid,-] (1.5,1)--(p3);
\draw[draw, solid,->] (p2)--(1.5,-1);
\draw[draw, solid,-] (1.5,-1)--(p3);
\draw[draw, solid,-<] (p3)--(4.5,0);
\draw[draw, solid,-] (4.5,0)--(p4);
\draw[draw, solid,->] (p3)--(4.5,-1);
\draw[draw, solid,-] (4.5,-1)--(p5);
\draw[draw, solid,-<] (p4)--(7.5,0);
\draw[draw, solid,-] (7.5,0)--(p6);
\draw[draw, solid,->] (p5)--(7.5,-1);
\draw[draw, solid,-] (7.5,-1)--(p6);
\draw[draw, solid,->] (p6)--(10.5,1);
\draw[draw, solid,-] (10.5,1)--(p7);
\draw[draw, solid,->] (p6)--(10.5,-1);
\draw[draw, solid,-] (10.5,-1)--(p8);
\end{tikzpicture}
\caption{The model arising from $S$-gluing two trinions along one puncture.}
\label{fig:Sgluingtrinions}
\end{figure}  

We first glue the two trinions along one puncture, say the one labeled by $L$, from each of them using the $S$-gluing. Recall that this gluing is done by gauging a diagonal combination of the $SU(2)$ symmetry of the puncture of each trinion\footnote{We note that here we don't need to turn on a CS term when gluing. This can be undestood by first gluing the trinions for $N_f=6$, where no CS term is needed, and then turning on the real mass deforations leading to lower $N_f$. Since the $SU(2)$ node does not see any flavors receiving the real masses, these do not generate a CS term.} and turning on the superpotential
\be\label{eq:Sgluingsuperpot}
\gd\mathcal{W}=M_L\cdot M_L'\,,
\ee
where $M_L$ is the moment map of one trinion and $M_L'$ that of the other one. Recalling \eqref{eq:momentmaps}, this translates into the sum of quartic and sextic couplings in the resulting model, which we depict in Figure \ref{fig:Sgluingtrinions}. We are not reporting them in the picture, but here there are many additional abelian symmetries compared to the original trinion model. As pointed out in \cite{Sacchi:2021afk,Sacchi:2021wvg} in the context of compactifications of $5d$ SCFTs on surfaces of genus one, one might need to turn on suitable monopole superpotential terms during the gluing on top of \eqref{eq:Sgluingsuperpot} in order to break some combinations of these symmetries. For the moment we will remain agnostic about it and we will reconsider this issue in the final genus 2 model. 

\begin{figure}[t]
\center
\begin{tikzpicture}[baseline=0, font=\scriptsize]
\node[draw, circle] (l) at (0,0) {\normalsize $3_{\frac{6-N_f}{2}}$};
\node[draw, circle] (r) at (6,0) {\normalsize $3_{\frac{N_f-6}{2}}$};
\node[draw, circle] (c1) at (3,0) {\large $\,\, 2 \,\,$};
\node[draw, circle] (c2) at (3,2) {\large $\,\, 2 \,\,$};
\node[draw, circle] (c3) at (3,4) {\large $\,\, 2 \,\,$};
\node[draw, rectangle] (b) at (3,-3) {\large $N_f-2$};

\draw[draw, solid,-<] (l)--(1.5,0);
\draw[draw, solid,-] (1.5,0)--(c1);
\draw[draw, solid,-<] (l)--(1.5,1);
\draw[draw, solid,-] (1.5,1)--(c2);
\draw[draw, solid,-<] (l)--(1.5,2);
\draw[draw, solid,-] (1.5,2)--(c3);
\draw[draw, solid,->] (r)--(4.5,0);
\draw[draw, solid,-] (4.5,0)--(c1);
\draw[draw, solid,->] (r)--(4.5,1);
\draw[draw, solid,-] (4.5,1)--(c2);
\draw[draw, solid,->] (r)--(4.5,2);
\draw[draw, solid,-] (4.5,2)--(c3);
\draw[draw, solid,->] (l)--(1.5,-1.5);
\draw[draw, solid,-] (1.5,-1.5)--(b);
\draw[draw, solid,-<] (r)--(4.5,-1.5);
\draw[draw, solid,-] (4.5,-1.5)--(b);

\node[] at (1.65,-0.3) {\normalsize $x^{\frac{1}{3}}u^6a^{-1}$};
\node[] at (1.9,-0.3+1) {\normalsize $x^{\frac{1}{3}}v^6a^{-1}$};
\node[] at (0.6,2) {\normalsize $x^{\frac{1}{3}}w^6a^{-1}$};
\node[] at (6-1.65,-0.3) {\normalsize $x^{\frac{1}{3}}u^{-6}ay$};
\node[] at (6-1.9,-0.3+1) {\normalsize $x^{\frac{1}{3}}v^{-6}ay$};
\node[] at (6-0.5,2) {\normalsize $x^{\frac{1}{3}}w^{-6}ay$};
\node[] at (-0.1,-1.5) {\normalsize $x^{\frac{2}{3}}u^{-2}v^{-2}w^{-2}a^3$};
\node[] at (6.1,-1.5) {\normalsize $x^{\frac{2}{3}}u^{2}v^{2}w^{2}a^{-3}y^{-1}$};
\end{tikzpicture}
\caption{The $3d$ $\mathcal{N}=2$ Lagrangian for the compactification of the $5d$ rank 1 $E_{N_f+1}$ SCFT on a genus 2 surface with no flux. The superpotential breaks one abelian symmetry corresponding to $y=1$.}
\label{fig:Genus2}
\end{figure}
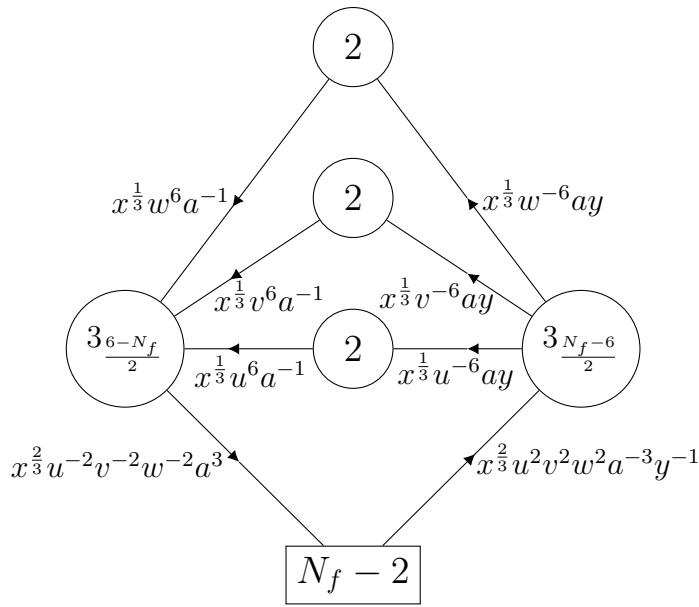

We then glue the remaining four punctures in pairs using the $S$-gluing again. The result is the model in Figure \ref{fig:Genus2}, which corresponds to the compactification of the $5d$ rank 1 $E_{N_f+1}$ SCFT on a genus 2 surface with no flux. Again remembering \eqref{eq:momentmaps} we can see that the superpotential for the perturbative matter part of the theory consists of quartic as well as sextic interactions. In particular the latter break a would be $U(1)$ global symmetry which we parameterized in Figure \ref{fig:Genus2} with the fugacity $y$. If we compute the index of this model for $N_f=6$, we find that this symmetry obstructs the expected global symmetry enhancement to $E_{7}$. Setting $y=1$ we can instead rewrite the index in terms of $E_7$ characters. Moreover, once this $U(1)$ symmetry is broken we can see that the fundamental, that is with magnetic flux 1, monopoles of the $SU(2)$ gauge nodes are exactly marginal, meaning that the charge assignment of the matter fields is compatible with having these monopoles turned on in the superpotential. We then assume that when performing the gluing one should also turn on such a monopole superpotential.

The index for $N_f=6$ after setting $y=1$ can be written in terms of $E_7$ characters as follows:
\be\label{ind_E7g2}
\mathcal{I}=1+(4+3+{\bf 133})x^2+\cdots\,.
\ee
The embedding of $E_7$ inside the manifest $SU(4)\times U(1)^4$ symmetry is
\bea
{\bf 56}&\to& \ga^2\gb^{-1}{\bf 6}+\ga^{-2}\gb\bar{\bf 6}+\ga^{-1}\gb^{-1}(\gc+\gc^{-1}){\bf 4}+\ga\gb(\gd+\gd^{-1}){\bf 4}+\ga\gb(\gc+\gc^{-1})\bar{\bf 4}+\nn\\
&+&\ga^{-1}\gb^{-1}(\gd+\gd^{-1})\bar{\bf 4}+\gb^3+\gb^{-3}+\ga^{-4}\gb^{-1}+\ga^4\gb+\nn\\
&+&(\ga^2\gb^{-1}+\ga^{-2}\gb)(\gc\gd+\gc\gd^{-1}+\gc^{-1}\gd+\gc^{-1}\gd^{-1})\,,
\eea
where we defined
\bea
&a=\ga^{\frac{1}{6}}\gb^{\frac{1}{6}}\gd^{\frac{1}{6}},\nn\\
&u=\ga^{\frac{1}{12}}\gb^{-\frac{1}{12}}\gc^{-\frac{1}{12}}\gd^{\frac{1}{18}},\nn\\
&v=\ga^{\frac{1}{12}}\gb^{-\frac{1}{12}}\gc^{\frac{1}{12}}\gd^{\frac{1}{18}},\nn\\
&w=\ga^{\frac{1}{4}}\gb^{\frac{1}{12}}\gd^{-\frac{1}{36}}\,.
\eea
This result is also consistent with the spectrum expected from $5d$ as we reviewed in Subsection \ref{subsec:tests}, since for genus $g=2$ and no flux we should have a contribution $3g-3+(g-1){\bf 133}=3+{\bf 133}$ to the order $x^2$ of the index. One discrepancy with the $5d$ expectations is that we have 4 additional marginal operators. This is due to the fact that once we set $y=1$ we have many more monopole operators that are exactly marginal than just the fundamental $SU(2)$ monopoles mentioned above. This deviation appears to be accidental as we shall soon show that it disappears once one goes to either higher genus or higher flux.

\begin{figure}[t]
\center
\begin{tikzpicture}[baseline=0, font=\scriptsize]
\node[draw, circle] (l) at (0,0) {\normalsize $3_{\frac{6-N_f}{2}}$};
\node[draw, circle] (r) at (6,0) {\normalsize $3_{\frac{N_f-6}{2}}$};
\node[draw, rectangle] (b) at (3,-3) {\large $N_f-2$};

\draw[draw, solid,-<] (l)--(3,0);
\draw[draw, solid,-] (3,0)--(r);
\draw[draw, solid,-<] (l) edge [out=30,in=180,loop,looseness=1] (3,1);
\draw[draw, solid,-] (3,1) edge [out=0,in=150,loop,looseness=1] (r);
\draw[draw, solid,-<] (l) edge [out=60,in=180,loop,looseness=1] (3,2);
\draw[draw, solid,-] (3,2) edge [out=0,in=120,loop,looseness=1] (r);
\draw[draw, solid,->] (l)--(1.5,-1.5);
\draw[draw, solid,-] (1.5,-1.5)--(b);
\draw[draw, solid,-<] (r)--(4.5,-1.5);
\draw[draw, solid,-] (4.5,-1.5)--(b);

\draw[black,solid,<->] (l) edge [out=70,in=110,loop,looseness=3]  (l);
\draw[black,solid,>-<] (r) edge [out=70,in=110,loop,looseness=3]  (r);
\draw[black,solid,<->] (l) edge [out=70+55,in=110+55,loop,looseness=3]  (l);
\draw[black,solid,>-<] (r) edge [out=70-55,in=110-55,loop,looseness=3]  (r);
\draw[black,solid,<->] (l) edge [out=70+2*55,in=110+2*55,loop,looseness=3]  (l);
\draw[black,solid,>-<] (r) edge [out=70-2*55,in=110-2*55,loop,looseness=3]  (r);

\node[] at (3,-0.3) {\normalsize $x^{\frac{2}{3}}$};
\node[] at (3,1-0.3) {\normalsize $x^{\frac{2}{3}}$};
\node[] at (3,2-0.3) {\normalsize $x^{\frac{2}{3}}$};
\node[] at (-0.1,-1.5) {\normalsize $x^{\frac{2}{3}}u^{-2}v^{-2}w^{-2}a^3$};
\node[] at (6.1,-1.5) {\normalsize $x^{\frac{2}{3}}u^{2}v^{2}w^{2}a^{-3}$};
\node[] at (-2.1,-0.3) {\normalsize $x^{\frac{2}{3}}u^{12}a^{-2}$};
\node[] at (-1.9,0.6) {\normalsize $x^{\frac{2}{3}}v^{12}a^{-2}$};
\node[] at (0,1.5) {\normalsize $x^{\frac{2}{3}}w^{12}a^{-2}$};
\node[] at (6+2.1,-0.3) {\normalsize $x^{\frac{2}{3}}u^{-12}a^{2}$};
\node[] at (6+2,0.6) {\normalsize $x^{\frac{2}{3}}v^{-12}a^{2}$};
\node[] at (6.1,1.5) {\normalsize $x^{\frac{2}{3}}w^{-12}a^{2}$};
\end{tikzpicture}
\caption{Dual description for the $3d$ $\mathcal{N}=2$ model arising from the compactification of the $5d$ rank 1 $E_{N_f+1}$ SCFT on a genus 2 surface with no flux. The arcs on the gauge nodes denote chiral fields in the antisymmetric representation or its complex conjugate, depending on the orientation of the arrows.}
\label{fig:Genus2dual}
\end{figure}
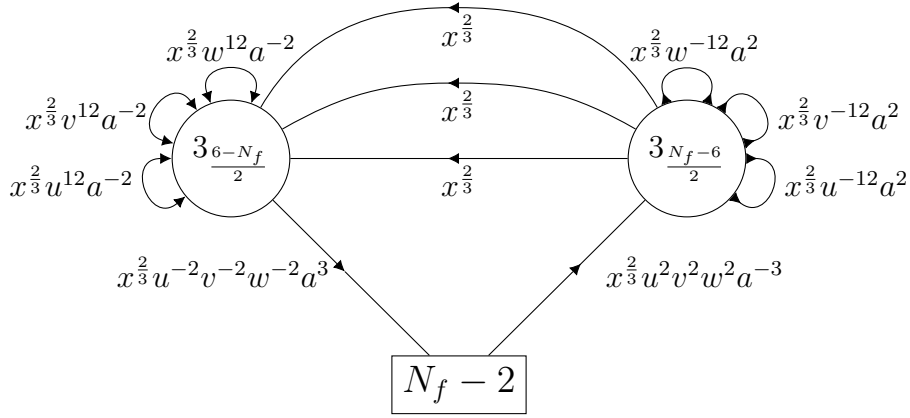 

The genus 2 model of Figure \ref{fig:Genus2} has a dual description in terms of a simpler quiver, similarly to what was done for the analogous $4d$ models in \cite{Razamat:2020bix}. Indeed, the $SU(2)$ nodes locally look like an $SU(2)$ gauge theory with 6 fundamental chirals and superpotential $\mathcal{W}=\mathfrak{M}$, where $\mathfrak{M}$ denotes the fundamental monopole, which is known to be dual to a WZ model of 15 chirals $M_{ij}$ with superpotential $\hat{\mathcal{W}}=\mathrm{Pf}(M)$ \cite{Aharony:2013dha}. If we apply this duality to each of the $SU(2)$ nodes of the quiver in Figure \ref{fig:Genus2} we remove them and the 15 fields split into an antisymmetric for each $SU(3)$ node and a bifundamental between them. The result is thus the theory in Figure \ref{fig:Genus2dual}, where now the perturbative superpotential consists of cubic interactions for each triangle made of the anti-fundamental for the left $SU(3)$, the fundamental for the right $SU(3)$ and one copy of the $SU(3)\times SU(3)$ bifundamentals and other cubic interactions involving two antisymmetrics, one from each $SU(3)$ gauge node, and one copy of the $SU(3)\times SU(3)$ bifundamentals. Notice that this is again necessary in order to kill one abelian symmetry. We find that the index of this theory coincides with \eqref{ind_E7g2}.  

\begin{figure}[t]
\center
\begin{tikzpicture}[baseline=0, font=\scriptsize]
\node[draw, circle] (l) at (0,0) {\normalsize $3_{\frac{6-N_f}{2}}$};
\node[draw, circle] (r) at (6,0) {\normalsize $3_{\frac{N_f-6}{2}}$};
\node[draw, rectangle] (b) at (3,-3) {\normalsize $N_f+1$};

\draw[draw, solid,-<] (l)--(3,0);
\draw[draw, solid,-] (3,0)--(r);
\draw[draw, solid,-<] (l) edge [out=30,in=180,loop,looseness=1] (3,1);
\draw[draw, solid,-] (3,1) edge [out=0,in=150,loop,looseness=1] (r);
\draw[draw, solid,-<] (l) edge [out=60,in=180,loop,looseness=1] (3,2);
\draw[draw, solid,-] (3,2) edge [out=0,in=120,loop,looseness=1] (r);
\draw[draw, solid,->] (l)--(1.5,-1.5);
\draw[draw, solid,-] (1.5,-1.5)--(b);
\draw[draw, solid,-<] (r)--(4.5,-1.5);
\draw[draw, solid,-] (4.5,-1.5)--(b);

\node[] at (3,-0.3) {\normalsize $x^{\frac{2}{3}}$};
\node[] at (3,1-0.3) {\normalsize $x^{\frac{2}{3}}$};
\node[] at (3,2-0.3) {\normalsize $x^{\frac{2}{3}}$};
\node[] at (-0.1,-1.5) {\normalsize $x^{\frac{2}{3}}d$};
\node[] at (6.1,-1.5) {\normalsize $x^{\frac{2}{3}}d^{-1}$};
\end{tikzpicture}
\caption{Equivalent representation of the dual description for the $3d$ $\mathcal{N}=2$ model arising from the compactification of the $5d$ rank 1 $E_{N_f+1}$ SCFT on a genus 2 surface with no flux.}
\label{fig:Genus2dualalt}
\end{figure}
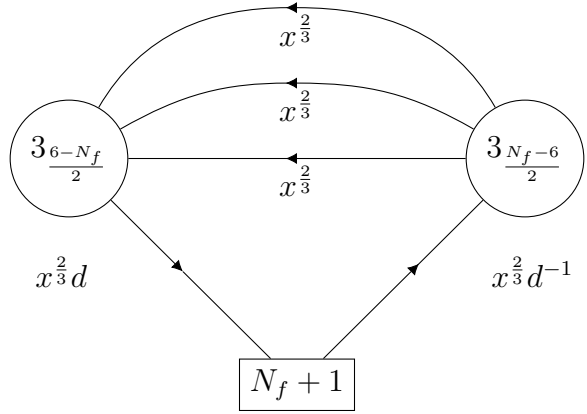

It is more convenient to work with the theory in Figure \ref{fig:Genus2dual} rather than the one in Figure \ref{fig:Genus2}, which is what we will do in the reminder of this subsection. One reason is that this has less gauge nodes, which makes computations easier. Moreover, part of the larger non-abelian symmetry can be seen more easily from the model in Figure \ref{fig:Genus2dual}. In fact, using that the complex conjugate of the antisymmetric representation of $SU(3)$ is actually the fundamental representation and viceversa for the antisymmetric representation, we can redraw the model in Figure \ref{fig:Genus2dual} as the one in Figure \ref{fig:Genus2dualalt}. We can see that now the manifest global symmetry is $SU(N_f+1)\times U(1)_d$, so there is only one abelian symmetry. This can be understood from the superpotential, which now takes the simpler form
\be\label{superpotgenus2}
\mathcal{W}=\sum_{i,j=1}^3P_i(P_j)^2+\sum_{i=1}^3Q_LP_iQ_R\,,
\ee
where $Q_{L/R}$ denotes the left/right fundamental fields and $P_i$ for $i=1,2,3$ the three bifundamental fields.
The global symmetry is expected to get enhanced to $E_{N_f+1}$ in the IR. Indeed, for $N_f=6$ the result \eqref{ind_E7g2} can be rewritten as
\bea
\label{ind_E7g2alt}
\mathcal{I}&=&1+(4+3+1+{\bf 48}_{SU(7)}+d^{-6}{\bf 7}_{SU(7)}+d^6\overline{\bf 7}_{SU(7)}+d^3{\bf 35}_{SU(7)}+d^{-3}\overline{\bf 35}_{SU(7)})x^2=\nn\\
&=&1+(4+3+{\bf 133}_{E_7})x^2+\cdots\,.
\eea

Another piece of evidence for this symmetry enhancement can be obtained by computing the central charges appearing in the two-point functions of the conserved currents of $U(1)_{d}$ and $SU(7)$. Since these two symmetries combine together in the IR to form the group $E_7$, their central charges are expected to be related to each other in accordance with the way these symmetries are embedded in $E_7$. In general, the central charge of a subgroup $H\subset G$ is related to that of $G$ by 
\begin{equation}
	C_{H}=I_{H\hookrightarrow G}C_{G}
\end{equation}
where $I_{H\hookrightarrow G}$ is the corresponding embedding index, which for an embedding
\begin{equation}
	\boldsymbol{R}_{G}\rightarrow\sum_{i}\boldsymbol{R}_{H}^{(i)}
\end{equation}
 is given by 
\begin{equation}
	I_{H\hookrightarrow G}=\frac{\sum_{i}T_{\boldsymbol{R}_{H}^{(i)}}}{T_{\boldsymbol{R}_{G}}}
\end{equation}
where $T_{\boldsymbol{R}}$ denotes the Dynkin index of the representation $\boldsymbol{R}$ and should be replaced by $q^2_i$ when $H$ is a $U(1)$ group. As a result, the ratio of the central charges of two subgroups $H_1$ and $H_2$ which combine to give a larger enhanced symmetry should be given by the ratio of the corresponding embedding indices
\begin{equation}
	\label{RC}
 \frac{C_{H_{1}}}{C_{H_{2}}}=\frac{I_{H_{1}\hookrightarrow G}}{I_{H_{2}\hookrightarrow G}}\,.
\end{equation}
This is then a necessary condition for the symmetry enhancement that we can verify. Specifically, if $U(1)_{d}$ and $SU(7)$ of the $N_f=6$ model indeed form an $E_7$, their central charges should be related as in \eqref{RC}. In order to check that, we numerically compute the central charges of $U(1)_{d}$ and of the Cartan $\textrm{diag}\left(1,0,0,0,0,0,-1\right)$ of $SU(7)$, denoted by $C$, from the real part of the free energy (see appendix B of \cite{Sacchi:2021afk} and references therein for more details). We obtain 
\begin{equation}
	C_{d}=31.76\,\,\,,\,\,\,C_{C}=2.01
\end{equation}
corresponding to the ratio 
\begin{equation}
	\label{ratioC}
	\frac{C_{d}}{C_{C}}=15.8,
\end{equation}
which indeed matches (within the accuracy of the computation) the expectations from the enhancement to $E_7$ since 
\begin{equation}
I_{U\left(1\right)_{d}\hookrightarrow E_{7}}=63\,,\,\,\,I_{U\left(1\right)_{C}\hookrightarrow SU\left(7\right)}=4\,,\,\,\,I_{SU\left(7\right)\hookrightarrow E_{7}}=1
\end{equation}
and therefore 
\begin{equation}
\frac{C_{d}}{C_{C}}=\frac{I_{U\left(1\right)_{d}\hookrightarrow E_{7}}}{I_{U\left(1\right)_{C}\hookrightarrow SU\left(7\right)}I_{SU\left(7\right)\hookrightarrow E_{7}}}=15.75.
\end{equation}

Using the simpler description of Figure \ref{fig:Genus2dualalt} we can also easily investigate cases with lower $N_f$. For $N_f=5$ we get the index 
\bea
\label{ind_E6g2alt}
\mathcal{I}&=&1+(9+3+1+{\bf 35}_{SU(6)}+d^{-6}+d^6+(d^3+d^{-3}){\bf 20}_{SU(6)})x^2=\nn\\
&=&1+(9+3+{\bf 78}_{E_6})x^2+\cdots
\eea
which supports the expected enhancement of $U(1)_{d}$ and $SU(6)$ to $E_6$. As before, to get further support we can also compute the central charges of these symmetries and compare their ratio to the one expected from the enhancement and the embedding. Denoting the Cartan $\textrm{diag}\left(1,0,0,0,0,-1\right)$ of $SU(6)$ by $C$ as in the previous case and computing numerically the central charges, we find 
\begin{equation}
C_{d}=15.2\,\,\,,\,\,\,C_{C}=1.7\,\,\,,\,\,\,\frac{C_{d}}{C_{C}}=8.9. 
\end{equation}
This matches, within the accuracy of the computation, the ratio expected from the enhancement which is given by 
\begin{equation}
\frac{C_{d}}{C_{C}}=\frac{I_{U\left(1\right)_{d}\hookrightarrow E_{6}}}{I_{U\left(1\right)_{C}\hookrightarrow SU\left(6\right)}I_{SU\left(6\right)\hookrightarrow E_{6}}}=9.
\end{equation}

Turning to the case $N_f=4$, we find the index 
\bea
\label{ind_E5g2alt}
\mathcal{I}&=&1+(16+3+1+{\bf 24}_{SU(5)}+d^3\overline{\bf 10}_{SU(5)}+d^{-3}{\bf 10}_{SU(5)})x^2=\nn\\
&=&1+(16+3+{\bf 45}_{SO(10)})x^2+\cdots
\eea
which demonstrates the enhancement of $U(1)_{d}$ and $SU(5)$ to $SO(10)$. The central charges are given in this case by 
\begin{equation}
C_{d}=10.385\,\,\,,\,\,\,C_{C}=1.846\,\,\,,\,\,\,\frac{C_{d}}{C_{C}}=5.6257
\end{equation}
which again matches the ratio expected from the enhancement
\begin{equation}
\frac{C_{d}}{C_{C}}=\frac{I_{U\left(1\right)_{d}\hookrightarrow SO(10)}}{I_{U\left(1\right)_{C}\hookrightarrow SU\left(5\right)}I_{SU\left(5\right)\hookrightarrow SO(10)}}=5.625.
\end{equation}

Continuing to the theory corresponding to $N_f=3$, we have the index 
\bea
\label{ind_E4g2alt}
\mathcal{I}&=&1+(25+3+1+{\bf 15}_{SU(4)}+d^3\overline{\bf 4}_{SU(4)}+d^{-3}{\bf 4}_{SU(4)})x^2=\nn\\
&=&1+(25+3+{\bf 24}_{SU(5)})x^2+\cdots
\eea
and the enhancement of $U(1)_{d}$ and $SU(4)$ to $SU(5)$. The central charges are given by 
\begin{equation}
C_{d}=7.5\,\,\,,\,\,\,C_{C}=2.1\,\,\,,\,\,\,\frac{C_{d}}{C_{C}}=3.57,
\end{equation}
in agreement with the ratio expected from the enhancement 
\begin{equation}
\frac{C_{d}}{C_{C}}=\frac{I_{U\left(1\right)_{d}\hookrightarrow SU\left(5\right)}}{I_{U\left(1\right)_{C}\hookrightarrow SU\left(4\right)}I_{SU\left(4\right)\hookrightarrow SU\left(5\right)}}=3.6.
\end{equation}

We next arrive to the case $N_f=2$, where the index is given by 
\bea
\label{ind_E3g2alt}
\mathcal{I}&=&1+(36+3+1+{\bf 8}_{SU(3)}+d^3+d^{-3})x^2=\nn\\
&=&1+(36+3+{\bf 8}_{SU(3)}+{\bf 3}_{SU(2)})x^2+\cdots\,,
\eea
signaling the enhancement of $U(1)_{d}$ to $SU(2)$. Since this enhancement does not involve two different groups combining together to form a single larger one, the computation of the central charges will not teach us something new in this case. 

Overall, we see that in all the cases above the expected symmetry enhancement from $SU(N_f+1)\times U(1)_d$ to $E_{N_f+1}$ indeed takes place. We also observe the presence of marginal operators coming from the $5d$ stress-energy tensor and conserved current multiplets, in accrdance with equation \eqref{Indexexp}. However, we also observe the presence of additional marginal operators for which we do not have a higher dimensional interpretation. These appear to be accidental to the genus $2$ and no flux case, as we shall see that they disappear for higher genus or once flux is turned on.

Let us also comment that another advantage of working with the quiver of Figure \ref{fig:Genus2dualalt} rather than the original one in Figure \ref{fig:Genus2} is that we can in principle consider also values of $N_f$ which are smaller than 2, even $N_f=-1$. . We will investigate this possibility in more detail in the rest of this section.

\subsubsection{Flowing to $N_f<2$}

We can try to flow from the theory with $N_f=2$ to the one with $N_f=1$ with a mass deformation. The natural choice is a mass deformation in the model of Figure \ref{fig:Genus2dualalt} for $N_f=2$ that breaks $SU(3)$ to $SU(2)$, which just amounts to considering the same model for $N_f=1$. Notice that such deformation makes sense only in the dualized genus 2 model and not in the original trinion, since it would break some of the symmetries of the punctures. The index of the resulting model is
\be\label{ind_E2g2}
\mathcal{I}=1+\left(54+t^2+t^{-2}\right)x^2+\cdots=1+\left(50+3+{\bf 3}\right)x^2+\cdots\,,
\ee
which is written in terms of characters of $E_2=SU(2)_{t}\times U(1)_{d}$. This symmetry is in fact already manifest in the Lagrangian description of the theory. This result is compatible with the theory being the result of the compactification of the $5d$ $E_2$ SCFT on a genus 2 surface with no flux.

We can go on and perform a further mass deformation so to reach the model with $N_f=0$. Here we have two choices: we can either perform a real mass deformation for $U(1)_{d}$ or for $SU(2)_{t}$. This is compatible with the fact that from the $E_2$ SCFT in $5d$ we can either flow to the $E_1$ or to the $\tilde{E}_1$ SCFT. Indeed, the mass deformation for $U(1)_{d}$ preserves $SU(2)_t$ which is the symmetry of the $E_1$ SCFT, while a mass deformation for $SU(2)_t$ preserves $U(1)_d\times SU(1)_{t}=U(1)_{d}$ which is the symmetry of the $\tilde{E}_1$ SCFT.

Let us start considering the mass deformation for $SU(2)_t$, which amounts to considering the model of Figure \ref{fig:Genus2dualalt} for $N_f=0$. The index of the resulting model is
\be
\mathcal{I}=1+68x^2+\cdots\,,
\ee
which is compatible with the fact that the $SU(2)$ symmetry has been broken. Another consistency check that this is indeed the compactification of the $\tilde{E}_1$ SCFT is that the model doesn't have a 1-form symmetry \cite{Gaiotto:2014kfa} because of the fundamental matter fields. Indeed, a difference between the $E_1$ and the $\tilde{E}_1$ SCFTs is that the former has a 1-form symmetry, while the latter has not \cite{Morrison:2020ool,Albertini:2020mdx}.

Let us now consider a real mass deformation for $U(1)_d$, which should lead to the compactification of the $E_1$ SCFT. Notice that if we perform such a deformation, all the fields charged under $SU(2)_t$ become massive, so we would also lose this symmetry, in contradiction with the $5d$ expectation. We can in principle avoid this by combining the real mass deformation with a Coulomb branch vev in such a way that we follow the theory to a vacuum where some of the fields charged under $SU(2)_t$ remain massless. Nevertheless, we don't have at this point a clear interpretation of what this additional Coulomb branch vev in the $3d$ theory should correspond to in $5d$ and hence we don't have a satisfactory understanding of the resulting model. We leave this issue for future investigations.

Interestingly, we can also access the genus 2 compactification without flux of the $E_0$ SCFT by just setting $N_f=-1$ in the model in Figure \ref{fig:Genus2dualalt}. Notice that this is consistent with the fact that the $E_0$ theory can be obtained as a mass deformation of the $\tilde{E}_1$ theory, which recall corresponds to the model in Figure \ref{fig:Genus2dualalt} for $N_f=0$.  The index of this model is
\be\label{ind_E0g2}
\mathcal{I}=1+84x^2+327x^4+\cdots\,.
\ee
Unfortunately at the moment the only test that we have that this is the correct theory corresponding to the compactification of the $E_0$ SCFT is that it has a $\mathbb{Z}_3$ one-form symmetry coming from the diagonal combination of the center symmetries of the two $SU(3)$ gauge groups that is not screened by the bifundamental matter fields. This agrees with the known $\mathbb{Z}_3$ one-form symmetry of the $5d$ $E_0$ SCFT \cite{Morrison:2020ool,Albertini:2020mdx}. In a future work \cite{wip} we will compute some discrete anomalies of this model, which could in principle be matched with some $5d$ expectation.

\subsection{Higher genus}

\begin{figure}[t]
\center
\begin{tikzpicture}[baseline=0, font=\scriptsize]
\node[draw, circle] (g1) at (0,0) {\normalsize $3_{\frac{N_f-6}{2}}$};
\node[draw, circle] (g2) at (7,0) {\normalsize $3_{\frac{6-N_f}{2}}$};
\node[draw, circle] (g4) at (0,-6) {\normalsize $3_{\frac{N_f-6}{2}}$};
\node[draw, circle] (g3) at (7,-6) {\normalsize $3_{\frac{6-N_f}{2}}$};
\node[draw, rectangle] (b) at (3.5,-3) {\normalsize $N_f+1$};

\draw[draw, double distance=1.2mm, solid,-{>[scale=2.5]}] (g1)--(3.5,0);
\draw[draw, double distance=1.2mm, solid,-] (3.5,0)--(g2);
\draw[draw, solid,-{<[scale=2]}] (g2)--(7,-3);
\draw[draw,, solid,-] (7,-3)--(g3);
\draw[draw, double distance=1.2mm, solid,-{>[scale=2.5]}] (g3)--(3.5,-6);
\draw[draw, double distance=1.2mm, solid,-] (3.5,-6)--(g4);
\draw[draw, solid,-{<[scale=2]}] (g4)--(0,-3);
\draw[draw, solid,-] (0,-3)--(g1);
\draw[draw, solid,-{<[scale=2]}] (g1)--(1.75,-1.5);
\draw[draw, solid,-] (1.75,-1.5)--(b);
\draw[draw, solid,-{>[scale=2]}] (g2)--(5.25,-1.5);
\draw[draw, solid,-] (5.25,-1.5)--(b);
\draw[draw, solid,-{<[scale=2]}] (g3)--(5.25,-4.5);
\draw[draw, solid,-] (5.25,-4.5)--(b);
\draw[draw, solid,-{>[scale=2]}] (g4)--(1.75,-4.5);
\draw[draw, solid,-] (1.75,-4.5)--(b);

\node[] at (3.5,0.5) {\normalsize $x^{\frac{2}{3}}$};
\node[] at (3.5,-6.5) {\normalsize $x^{\frac{2}{3}}$};
\node[] at (-0.5,-3) {\normalsize $x^{\frac{2}{3}}$};
\node[] at (7.5,-3) {\normalsize $x^{\frac{2}{3}}$};
\node[] at (4.8,-1.2) {\normalsize $x^{\frac{2}{3}}d$};
\node[] at (4.8,-4.9) {\normalsize $x^{\frac{2}{3}}d^{-1}$};
\node[] at (2.5,-1.2) {\normalsize $x^{\frac{2}{3}}d^{-1}$};
\node[] at (2,-4.9) {\normalsize $x^{\frac{2}{3}}d$};
\end{tikzpicture}
\caption{The $3d$ $\mathcal{N}=2$ model arising from the compactification of the $5d$ rank 1 $E_{N_f+1}$ SCFT on a genus 3 surface with no flux. Double lines denote pairs of chiral fields in the same representations.}
\label{fig:g3f0}
\end{figure}
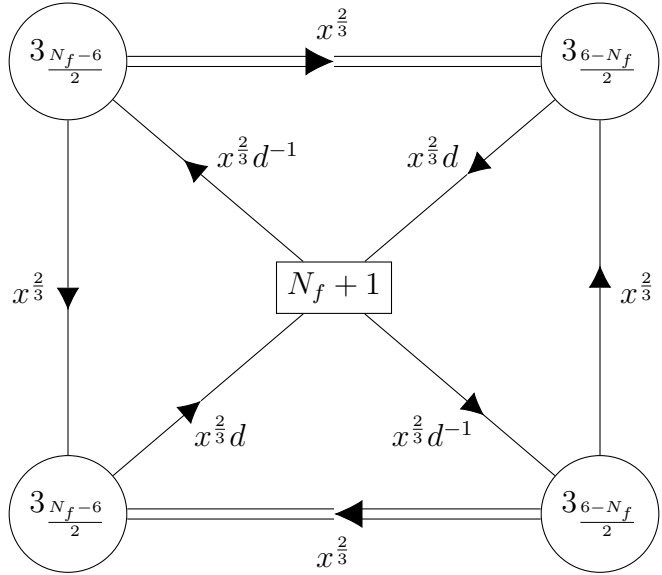 

In order to further validate our construction, we can use the trinion in Figure \ref{fig:3dtrinion} to construct the theory corresponding to the compactification on a Riemann surface of higher genus. Gluing four trinions with the $S$-gluing we can get a surface with genus $g=3$ and no flux, which we depict in Figure \ref{fig:g3f0}. This specific quiver is obtained after performing the gluings and dualizing all the $SU(2)$ gauge nodes, similarly to what we did for genus $g=2$. We have superpotential interactions that are analogous to those we had for $g=2$ and which determine the charge assignment summarized in the figure. In the general case of genus $g$ one needs to glue $2g-2$ trinions and so the model takes a form similar to the quiver in Figure \ref{fig:g3f0} but with $2g-2$ gauge nodes $SU(3)$.

Computing the index for $N_f=6$ we get
\bea
\label{ind_E7g3}
\mathcal{I}&=&1+(6+2(1+{\bf 48}_{SU(7)}+d^{-6}{\bf 7}_{SU(7)}+d^6\overline{\bf 7}_{SU(7)}+d^3{\bf 35}_{SU(7)}+d^{-3}\overline{\bf 35}_{SU(7)}))x^2=\nn\\
&=&1+(6+2\times{\bf 133}_{E_7})x^2+\cdots\,.
\eea
We can see that it forms representations of the expected enhanced $E_7$ symmetry with the same embedding of the symmetry $SU(7)\times U(1)$ that is manifest in the quiver of Figure \ref{fig:g3f0} that we also had for genus $g=2$. Moreover, the operators appearing at order $x^2$ agree with the $5d$ expectation, in particular they have the correct multiplicity that they should have for genus $g=3$, namely we have $3g-3=6$ marginal operators uncharged under all the symmetries and $g-1+qF=2$ operators in the $\bf 133$ of $E_7$. Notice that in this case, unlike what we saw for genus $g=2$, there are no extra marginal operators.

We can similarly check that the index is compatible with our expectations also for lower values of $N_f$. For $N_f=5$, we get
\bea
\label{ind_E6g3}
\mathcal{I}&=&1+(6+2(1+{\bf 35}_{SU(6)}+d^{-6}+d^6+(d^3+d^{-3}){\bf 20}_{SU(6)}))x^2=\nn\\
&=&1+(6+2\times{\bf 78}_{E_6})x^2+\cdots\,.
\eea
For $N_f=4$ we get
\bea
\label{ind_E5g3}
\mathcal{I}&=&1+(6+2(1+{\bf 24}_{SU(5)}+d^3\overline{\bf 10}_{SU(5)}+d^{-3}{\bf 10}_{SU(5)}))x^2=\nn\\
&=&1+(6+2\times{\bf 45}_{SO(10)})x^2+\cdots\,.
\eea
For $N_f=3$ we get
\bea
\label{ind_E4g2alt}
\mathcal{I}&=&1+(6+2(1+{\bf 15}_{SU(4)}+d^3\overline{\bf 4}_{SU(4)}+d^{-3}{\bf 4}_{SU(4)}))x^2=\nn\\
&=&1+(6+2\times{\bf 24}_{SU(5)})x^2+\cdots\,.
\eea
For $N_f=2$ we get
\bea
\label{ind_E3g2alt}
\mathcal{I}&=&1+(6+2(1+{\bf 8}_{SU(3)}+d^3+d^{-3}))x^2=\nn\\
&=&1+(6+2({\bf 8}_{SU(3)}+{\bf 3}_{SU(2)}))x^2+\cdots\,.
\eea
As for genus $2$, we can in principal go also to cases with $N_f<2$, but as the index computation for these proves to be more demanding, we shall content ourselves with the $N_f\geq 2$ cases.

In all of these cases we can see that the index is compatible with the expected enhancement of the manifest $SU(N_f+1)\times U(1)_d$ symmetry to $E_{N_f+1}$ and that the spectrum of operators is compatible with the $5d$ expectation.

\section{Compactifications on genus two surfaces with flux}
\label{sec:flux}

\begin{figure}[t]
\center
\begin{tikzpicture}[baseline=0, font=\scriptsize,scale=1.15]
\node[draw, circle] (g1) at (0,0) {\small $3_{k}$};
\node[draw, circle] (g2) at (12,0) {\small $3_{k}$};
\node[draw, circle] (g3) at (3,-4) {\small $2_{\text{-}k}$};
\node[draw, circle] (g4) at (9,-4) {\small $2_{\text{-}k}$};
\node[draw, circle] (g5) at (6,-6) {\small $2_{\text{-}k}$};
\node[draw] (f) at (6,0) {\small $N_f-2$};

\draw[draw, solid,<->] (g3)--(g5);
\draw[draw, solid,<->] (g3)--(g4);
\draw[draw, solid,<->] (g4)--(g5);
\draw[draw, solid,->] (g1)--(3,0);
\draw[draw, solid,-] (3,0)--(f);
\draw[draw, solid,->] (g2)--(9,0);
\draw[draw, solid,-] (9,0)--(f);
\draw[draw, solid,->] (g3)--(1.5,-2);
\draw[draw, solid,-] (1.5,-2)--(g1);
\draw[draw, solid,->] (g4)--(4.5,-2);
\draw[draw, solid,-] (4.5,-2)--(g1);
\draw[draw, solid,->] (g5)--(3,-3);
\draw[draw, solid,-] (3,-3)--(g1);
\draw[draw, solid,->] (g3)--(7.5,-2);
\draw[draw, solid,-] (7.5,-2)--(g2);
\draw[draw, solid,->] (g4)--(10.5,-2);
\draw[draw, solid,-] (10.5,-2)--(g2);
\draw[draw, solid,->] (g5)--(9,-3);
\draw[draw, solid,-] (9,-3)--(g2);
\draw[draw, solid,->] (f)--(5,-4/3);
\draw[draw, solid,-] (5,-4/3)--(g3);
\draw[draw, solid,->] (f)--(7,-4/3);
\draw[draw, solid,-] (7,-4/3)--(g4);
\draw[draw, solid,->] (f)--(6,-4/3);
\draw[draw, solid,-] (6,-4/3)--(g5);

\node[] at (3,0.4) {\footnotesize $x^{\frac{2}{3}}a^3u^{-2}v^{-2}w^{-2}$};
\node[] at (9,0.4) {\footnotesize $x^{\frac{2}{3}}a^3u^{-2}v^{-2}w^{-2}$};
\node[] at (0.7,-2) {\footnotesize $x^{\frac{1}{3}}a^{-1}u^{6}$};
\node[] at (2.8,-2) {\footnotesize $x^{\frac{1}{3}}a^{-1}v^{6}$};
\node[] at (2.6,-0.6) {\footnotesize $x^{\frac{1}{3}}a^{-1}w^{6}$};
\node[] at (12-0.6,-2) {\footnotesize $x^{\frac{1}{3}}a^{-1}w^{6}$};
\node[] at (12-2.8,-2) {\footnotesize $x^{\frac{1}{3}}a^{-1}v^{6}$};
\node[] at (12-2.4,-0.6) {\footnotesize $x^{\frac{1}{3}}a^{-1}u^{6}$};
\node[] at (4.6,-0.8) {\footnotesize $xa^{-2}u^{-4}$};
\node[] at (4.6,-1.2) {\footnotesize $v^2w^{2}$};
\node[] at (5.4,-1.5) {\footnotesize $xa^{-2}$};
\node[] at (5.4,-1.9) {\footnotesize $u^{2}v^{-4}w^{2}$};
\node[] at (7.5,-0.8) {\footnotesize $xa^{-2}u^{2}$};
\node[] at (7.5,-1.2) {\footnotesize $v^2w^{-4}$};

\node[] at (4.5,-5.8) {\footnotesize $xa^3u^{-6}v^{-12}$};
\node[] at (7.55,-5.8) {\footnotesize $xa^3v^{-12}w^{-6}$};
\node[] at (3,-4.7) {\footnotesize $xa^3u^{-12}v^{-6}$};
\node[] at (9,-4.7) {\footnotesize $xa^3v^{-6}w^{-12}$};
\node[] at (4.85,-3.8) {\footnotesize $xa^3u^{-12}w^{-6}$};
\node[] at (9-1.7,-3.8) {\footnotesize $xa^3u^{-6}w^{-12}$};

\end{tikzpicture}
\caption{The $3d$ $\mathcal{N}=2$ Lagrangian for the compactification of the $5d$ rank 1 $E_{N_f+1}$ SCFT on a genus 2 surface with one unit of flux. The Chern--Simons levels are written in terms of $k=\frac{6-N_f}{2}$. The lines with double arrows between pairs of $SU(2)$ nodes stand for a pair of chirals, one of which is a doublet of one node and a singlet of the other node and the other chiral the opposite. The charges of each of these chirals is written close to the arrow that points to the node under which it is a doublet.}
\label{Genus2Phi}
\end{figure}
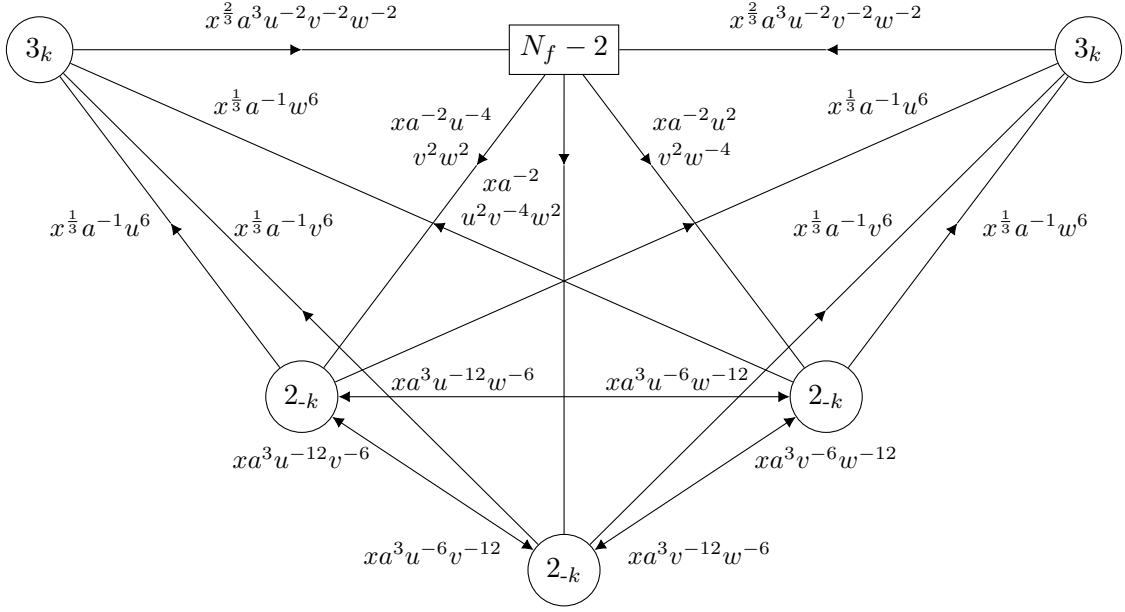

We can also try to construct a genus 2 surface via $\Phi$-gluing. As we reviewed in Subsection \ref{subsec:gluing}, while in the $S$-gluing we turned on the superpotential interaction \eqref{eq:Sgluingsuperpot} between the moment maps $M_L$ and $M_L'$ of the glued punctures of the two trinions, in the $\Phi$-gluing we introduce $N_f$ chirals $\Phi$ in the fundamental representation of the gauged $SU(2)$ puncture symmetry with the interaction
\be\label{eq:Phigluingsuperpot}
\gd\mathcal{W}=\Phi\cdot\left(M_L + M_L'\right)\,.
\ee
The resulting model is depicted in Figure \ref{Genus2Phi}. The full superpotential consists of various types of cubic and quartic couplings, associated with various closed loops in the quiver diagram:
\begin{itemize}
\item cubic interactions involving an $SU(3)\times SU(N_f-2)$ bifundamental, an $SU(N_f-2)\times SU(2)$ bifundamental and an $SU(2)\times SU(3)$ bifundamental;
\item quartic interactions involving one of the chirals which is a doublet of one $SU(2)$ and a singlet of another $SU(2)$ (represented by a line with a double arrow in Figure \ref{Genus2Phi}), an $SU(3)\times SU(2)$ bifundamental for the former $SU(2)$ and the square of an $SU(3)\times SU(2)$ bifundamental for the latter $SU(2)$ (so to make an $SU(2)$ singlet).
\end{itemize}
These superpotential interactions constrain the manifest global symmetry of the model to be $SU(N_f-2)\times U(1)^4$. Notice that the rank is $N_f+1$, which matches that of the $5d$ $E_{N_f+1}$ symmetry. This suggests that we do not need to consider any additional superpotential term involving monopole operators to break additional unwanted symmetries. We will indeed see momentarily that the model without monopole superpotential passes all the tests required by compatibility with the compactification picture.

When preforming a $\Phi$-gluing, the fluxes of the glued trinions add up so that, after gluing three pairs of punctures, we get a genus 2 surface with a flux which is twice that of the single trinion. In the $U(1)\times SO(2N_f)\subset E_{N_f+1}$ basis, this is given by $(\frac{\sqrt{8-N_f}}{2};\underbrace{\frac{1}{2},\frac{1}{2},...,\frac{1}{2}}_{N_f})$.

Let us start analyzing the case $N_f=6$. Here the model corresponds to a unit of flux for a $U(1)$ inside $E_7$ whose commutant is $SO(12)$. The index reads
\bea
\mathcal{I}&=&1+(7+{\bf 15}+\gb^{\pm2}\gc^{\pm1}\gd^{\pm1}+\gc^{\pm2}+\gc^{\pm2}+{\bf 6}(\gb^{\pm2}+\gc^{\pm1}\gd^{\pm1}))+\nn\\
&+&3\ga^2+3\ga^{-2}+2\ga({\bf 4}(\gb\gc^{\pm1}+\gb^{-1}\gd^{\pm1})+\overline{\bf 4}(\gb^{-1}\gc^{\pm1}+\gb\gd^{\pm1}))x^2+\cdots\nn\\
&=&1+(4+{\bf 66}_{SO(12)}+3\ga^2+3\ga^{-2}+2\ga{\bf 32}_{SO(12)})x^2+\cdots\,,
\eea
where we redefined
\bea
&a=\gb^{-\frac{1}{6}}\gc^{-\frac{1}{6}},\nn\\
&u=\ga^{-\frac{1}{18}}\gb^{-\frac{5}{36}}\gc^{\frac{1}{36}},\nn\\
&v=\ga^{-\frac{1}{18}}\gb^{\frac{1}{36}}\gc^{-\frac{1}{18}}\gd^{-\frac{1}{12}},\nn\\
&w=\ga^{-\frac{1}{18}}\gb^{\frac{1}{36}}\gc^{-\frac{1}{18}}\gd^{\frac{1}{12}}\,.
\eea
The $SO(12)\times U(1)$ symmetry of the model is precisely the subgroup of the $5d$ $E_7$ symmetry preserved by the flux. The embedding of the former inside the latter is
\bea\label{eq:E7decomp}
{\bf 133}\to{\bf 1}^0\oplus {\bf 66}^0\oplus {\bf 1}^{\pm2}\oplus {\bf 32}^{\pm1}\,.
\eea
The spectrum that we see in the index of the $3d$ theory is also consistent with the $5d$ expectation. Indeed, in accordance with \eqref{Indexexp}, each state in the decomposition \eqref{eq:E7decomp} of the $5d$ flavor symmetry current appears with coefficient $g-1+qF$, where $g$ is the genus, $q$ is the charge of the state under the $U(1)$ symmetry and $F$ is the flux. Moreover, one also expects $3g-3$ operators uncharged under all the symmetries, which come from the $5d$ stress-energy tensor. The only exception is the state ${\bf 1}^{-2}$ which appears in the index as $3\ga^{-2}$ while from $5d$ we would expect a coefficient $g-1+qF=2-1-2=-1$. We expect this to be due to extra operators that appear in the spectrum of the $3d$ theory, which generically occurs for low genus and flux. 

Let us consider now the case of $N_f=5$. This is obtained from the previous model via a mass deformation, which induces CS levels $\frac{1}{2}$ for the $SU(3)$ gauge nodes and $-\frac{1}{2}$ for the $SU(2)$ ones, see Figure \ref{Genus2Phi}. We expect this model to be associated with a unit of flux breaking the $5d$ $E_6$ global symmetry down to $SU(6)\times U(1)$. This should get enhanced from the manifest $SU(3)\times U(1)^4$ global symmetry of the model. Indeed, we can write the index in terms of characters for this symmetry by performing the following redefinition of the fugacities:
\bea
&a=\gb^{-\frac{1}{2}},\nn\\
&u=\ga^{-\frac{1}{18}}\gb^{-\frac{1}{12}}\gc^{-\frac{1}{6}},\nn\\
&v=\ga^{-\frac{1}{18}}\gb^{-\frac{1}{12}}\gc^{\frac{1}{12}}\gd^{-\frac{1}{12}},\nn\\
&w=\ga^{-\frac{1}{18}}\gb^{-\frac{1}{12}}\gc^{\frac{1}{12}}\gd^{\frac{1}{12}}\,.
\eea
With this new parametrization, the index reads
\bea
\mathcal{I}&=&1+\left(7+{\bf 8}+\gc^{\pm3}\gd^{\pm1}+\gd^{\pm2}+\gb^2{\bf 3}\left(\gc^{-1}\gd^{\pm1}+\gc^2\right)+\gb^{-2}\overline{\bf 3}\left(\gc\gd^{\pm1}+\gc^{-2}\right)\right.\nn\\
&+&\left.3\ga^2+8\alpha^{-2}+2\ga\left(\gb^3+\frac{1}{\gb^3}+\gb^{-1}{\bf 3}\left(\gc^{-1}\gd^{\pm1}+\gc^2\right)+\gb\overline{\bf 3}\left(\gc\gd^{\pm1}+\gc^{-2}\right)\right)\right)x^2+\cdots\nn\\
&=&1+\left(4+{\bf 35}_{SU(6)}+3\ga^2+8\alpha^{-2}+2\ga{\bf 20}_{SU(6)}\right)x^2+\cdots\,.\nn\\
\eea
The $SU(6)\times U(1)$ symmetry of the model is precisely the subgroup of the $5d$ $E_6$ symmetry preserved by the flux. The embedding of the former inside the latter is
\bea\label{eq:E6decomp}
{\bf 78}\to{\bf 1}^0\oplus {\bf 35}^0\oplus {\bf 1}^{\pm2}\oplus {\bf 20}^{\pm1}\,.
\eea
The spectrum that we see in the index of the $3d$ theory is also consistent with the $5d$ expectation, again except for the state ${\bf 1}^{-2}$ whose contribution to the $3d$ index has the wrong coefficient due to extra operators.

For $N_f=4$ the flux preserves the $SU(4)\times SU(2)\times U(1)$ subgroup of the $5d$ $E_5=SO(10)$ symmetry. The former should be enhanced from the manifest $SU(2)\times U(1)^4$ symmetry of the model. Indeed, we can write the index in terms of characters for this symmetry by performing the following redefinition of the fugacities:
\bea
&a=\gb^{-\frac{1}{3}}\gc^{-\frac{1}{3}},\nn\\
&u=\ga^{-\frac{1}{18}}\gb^{-\frac{1}{9}}\gc^{-\frac{1}{36}}\gd^{\frac{1}{12}},\nn\\
&v=\ga^{-\frac{1}{18}}\gb^{\frac{1}{18}}\gc^{-\frac{1}{9}},\nn\\
&w=\ga^{-\frac{1}{18}}\gb^{-\frac{1}{9}}\gc^{-\frac{1}{36}}\gd^{-\frac{1}{12}}\,.
\eea
With this new parameterization, the index reads
\bea
\mathcal{I}&=&1+\left(7+{\bf 3}+\gc^2 + \gc^{-2}+\gd^2 + \gd^{-2}+(\gb^2+\gb^{-2})(\gc+\gc^{-1})(\gd+\gd^{-1})\right.\nn\\
&+&\left.3\ga^2+15\alpha^{-2}+2\ga\left(\gb^2+\gb^{-2}+\gc\gd+\gc^{-1}\gd+\gc\gd^{-1}+\gc^{-1}\gd^{-1}\right){\bf 2}\right)x^2+\cdots\nn\\
&=&1+\left(4+({\bf 15},{\bf 1})+({\bf 1},{\bf 3})+3\ga^2+15\alpha^{-2}+2\ga({\bf 6},{\bf 2})\right)x^2+\cdots\,.
\eea
The $SU(4)\times SU(2)\times U(1)$ symmetry of the model is precisely the subgroup of the $5d$ $E_5=SO(10)$ symmetry preserved by the flux. The embedding of the former inside the latter is
\bea\label{eq:E5decomp}
{\bf 45}\to({\bf 1},{\bf 1})^0\oplus ({\bf 15},{\bf 1})^0\oplus ({\bf 1},{\bf 3})^0\oplus ({\bf 1},{\bf 1})^{\pm2}\oplus ({\bf 6},{\bf 2})^{\pm1}\,.
\eea
The spectrum that we see in the index of the $3d$ theory is also consistent with the $5d$ expectation, again except for the state $({\bf 1},{\bf 1})^{-2}$ whose contribution to the $3d$ index has the wrong coefficient due to extra operators.

For $N_f=3$ the flux preserves the $SU(3)\times U(1)^2$ subgroup of the $5d$ $E_4=SU(5)$ symmetry. The former should be enhanced from the manifest $U(1)^4$ symmetry of the model. Indeed, we can write the index in terms of characters for this symmetry by performing the following redefinition of the fugacities:
\bea
&a=\gb^{\frac{1}{2}},\nn\\
&u=\ga^{-\frac{1}{18}}\gb^{\frac{1}{12}}\gc^{\frac{1}{12}}\gd^{\frac{1}{12}},\nn\\
&v=\ga^{-\frac{1}{18}}\gb^{\frac{1}{12}}\gc^{\frac{1}{12}}\gd^{-\frac{1}{12}},\nn\\
&w=\ga^{-\frac{1}{18}}\gb^{\frac{1}{12}}\gc^{\frac{1}{12}}\,.
\eea
With this new parameterization, the index reads
\bea
\mathcal{I}&=&1+\left(7+(\gc^3+\gc^{-3})(\gd+\gd^{-1})+\gd^2+\gd^{-2}\right.\nn\\
&+&\left.3\ga^2+24\alpha^{-2}+2\ga\left(\gb\gc^2+\gb^{-1}\gc^{-2}+(\gb^{-1}\gc+\gb\gc^{-1})(\gd+\gd^{-1})\right)\right)x^2+\cdots\nn\\
&=&1+\left(5+{\bf 8}+3\ga^2+24\ga^{-2}2\ga(\gb^{-1}{\bf 3}+\gb\overline{\bf 3})\right)x^2+\cdots\,.
\eea
The $SU(3)\times U(1)^2$ symmetry of the model is precisely the subgroup of the $5d$ $E_4=SU(5)$ symmetry preserved by the flux. The embedding of the former inside the latter is
\bea\label{eq:E4decomp}
{\bf 24}\to2\times{\bf 1}^{(0,0)}\oplus {\bf 8}^{(0,0)}+{\bf 1}^{\pm2,0}\oplus{\bf 3}^{\pm1,-1}\oplus\overline{\bf 3}^{\pm1,1}\,.
\eea
The spectrum that we see in the index of the $3d$ theory is also consistent with the $5d$ expectation, again except for the state ${\bf 1}^{(-2,0)}$ whose contribution to the $3d$ index has the wrong coefficient due to extra operators.

Like in the previous cases, the computational complexity increases with decreasing flavors, and given the complicated quiver involved in this case, we shall content ourselves with the cases of $N_f\geq 3$.  

\section{Discussion}

In this paper we studied the compactifications of the $5d$ rank 1 Seiberg $E_{N_f+1}$ SCFTs on Riemann surfaces with genus which is higher than 1, generalizing the previous work  \cite{Sacchi:2021afk} in which the compactification surface was a torus. In order to do this, we began with finding the $3d$ model associated with the compactification on a trinion, which serves as a basic building block from which Riemann surfaces with any higher genus can be constructed. The $3d$ trinion model, in turn, was conjectured in two different ways which led to the same result. In the first, building on the knowledge of the $4d$ trinion model corresponding to the compactification of the $6d$ rank 1 E-string theory and the relation between the $6d/4d$ and $5d/3d$ setups via circle compactification, we obtained the $3d$ trinion model from the $4d$ one using circle reduction and a mass deformation. In the second approach, following the methodology of \cite{Razamat:2019mdt,Razamat:2020bix}, the same trinion model was obtained from the models discussed in \cite{Sacchi:2021afk,Sacchi:2021wvg} by a suitable deformation. 

After finding the trinion model, we turned to the study of theories corresponding to compactifying the rank 1 Seiberg $E_{N_f+1}$ SCFTs on Riemann surfaces with higher genus, obtained by gluing multiple trinions together. In particular, we checked that properties expected from this higher-dimensional origin, such as the existence of the correct global symmetries (which are usually enhanced in the IR) and the presence of specific operators descending from $5d$ multiplets, can indeed be identified. Since these properties are not manifest in the $3d$ UV Lagrangian description, checking them is a highly non-trivial test for the validity of our conjectured trinion model and the gluing procedure. We started by exploring the theories obtained by $S$-gluing trinions, corresponding to compactifications on Riemann surfaces with no flux for the global symmetry. Focusing on theories corresponding to genus 2 and 3, we employed the superconformal index and the central charges appearing in two-point functions of global symmetry currents to argue that the IR symmetry of these theories (which is usually enhanced compared to the UV one) is exactly the one expected from the $5d$ picture. Moreover, we observed that the index contains the contributions of the operators descending from special $5d$ multiplets. Then, in a similar way we analyzed the genus 2 theories obtained by $\Phi$-gluing trinions, corresponding to compactifications on Riemann surfaces with flux, and obtained perfect agreement with the $5d$ expectations. 

Several directions appear natural to be explored based on the results of this paper. First, here we focused on a specific family of $5d$ SCFTs, namely the rank 1 Seiberg theories with $E_{N_f+1}$ symmetry, and it will be interesting to generalize this analysis to more general theories and find their $3d$ trinion models. Another very interesting direction is to try to match 't Hooft anomalies between the $5d$ and $3d$ theories. In the $6d/4d$ setup, there are continues symmetries whose anomaly matching upon compactification proved to be extremely useful in both conjecturing and checking the $4d$ theory resulting from compactifying a given $6d$ SCFT. In the $5d/3d$ setup, even though there are no anomalies for continues symmetries, there are anomalies involving discrete symmetries. Specifically, certain anomalies of $5d$ SCFTs are known \cite{Morrison:2020ool,Albertini:2020mdx,Bhardwaj:2020phs,BenettiGenolini:2020doj,Apruzzi:2021nmk,Genolini:2022mpi} and we can try to match their dimensional reduction with those computed directly in the $3d$ models, along the lines of the field theory analysis of \cite{Tachikawa:2019dvq,Bergman:2020ifi,Beratto:2021xmn,Genolini:2022mpi,Bhardwaj:2022dyt,Mekareeya:2022spm,Bhardwaj:2023zix}.\footnote{Discrete anomalies of certain $3d$ theories have been computed also holographically, see for example \cite{Bergman:2020ifi,vanBeest:2022fss}. Moreover we point out that generalized symmetries of $3d$ $\mathcal{N}=4$ theories that are magnetic quivers of $5d$ theories have been studied for example in \cite{Closset:2020afy,Bhardwaj:2023zix,Nawata:2023rdx}.} This would provide an additional check of the proposals both of this paper and of \cite{Sacchi:2021afk,Sacchi:2021wvg}, and possibly have new predictions for the anomalies of certain $5d$ and $3d$ theories. We will address this topic of discrete anomaly matching in the $5d/3d$ setup in a following paper \cite{wip}.

\section*{Acknowledgments}
We would like to thank Shlomo Razamat for useful discussions. MS is partially supported by the ERC Consolidator Grant \#864828 “Algebraic Foundations of Supersymmetric Quantum Field Theory (SCFTAlg)” and by the Simons Collaboration for the Nonperturbative Bootstrap under grant \#494786 from the Simons Foundation. GZ is also partially supported by the Simons Foundation grant 815892. OS is supported by the Mani L. Bhaumik Institute for Theoretical Physics at UCLA. 

\bibliographystyle{JHEP}
\bibliography{ref}

\end{document}